\documentclass[aps,twocolumn,floatfix,groupedaddress,nofootinbib]{revtex4}
\usepackage{csquotes}
\usepackage{graphicx,color}
\usepackage{float}
\usepackage[caption=false]{subfig}
\usepackage{amsmath}
\usepackage{amssymb}
\usepackage{braket}
\usepackage{multirow}
\usepackage{dsfont}
\expandafter\let\csname equation*\endcsname\relax
\expandafter\let\csname endequation*\endcsname\relax

\begin{document}

\title{Integral equation theory based dielectric scheme for strongly coupled electron liquids\vspace*{-2.40mm}}
\author{P. Tolias$^{1}$, F. Lucco Castello$^{1}$ and T. Dornheim$^{2,3}$}
\affiliation{$^1$ Space and Plasma Physics - Royal Institute of Technology (KTH), SE-10044 Stockholm, Sweden\vspace*{-0.45mm}\\
             $^2$ Center for Advanced Systems Understanding (CASUS), D-02826 G\"orlitz, Germany\vspace*{-0.45mm}\\
             $^3$ Helmholtz-Zentrum Dresden-Rossendorf (HZDR), D-01328 Dresden, Germany\vspace*{-2.40mm}}
\begin{abstract}
\noindent In a recent paper, Lucco Castello \emph{et al.} [arXiv:2107.03537] provided an accurate parametrization of classical one-component plasma bridge functions that was embedded in a novel dielectric scheme for strongly coupled electron liquids. Here, this approach is rigorously formulated, its set of equations is formally derived and its numerical algorithm is scrutinized. Systematic comparison with available and new path integral Monte Carlo simulations reveals a rather unprecedented agreement especially in terms of the interaction energy and the long wavelength limit of the static local field correction.
\end{abstract}
\maketitle

\section{Introduction}\label{intro}

\noindent The quantum one-component plasma or jellium (referred to also as uniform electron gas or homogeneous electron gas) is universally acclaimed as one of the most important idealized systems in condensed matter physics\,\cite{intro01,intro02}, statistical mechanics\,\cite{intro03,intro04,intro05,intro06} and quantum chemistry\,\cite{intro07,intro08}. The jellium can serve as a model for many metals, under the simplifying assumption that the charge density of the positive ionic background is uniformly smeared out\,\cite{intro02,intro03,intro04,intro05,intro06}. Hence, initial investigations focused on its ground state at metallic densities\,\cite{intro09}.\,The ground state studies have led to remarkable insights such as the BCS superconductivity theory\,\cite{intro10}, Landau's Fermi-liquid theory\,\cite{intro11}, the Bohm-Pines quasi-particle picture of collective excitations\,\cite{intro12}, Wigner's crystallization paradigm\,\cite{intro13} and even the Kohn -Sham formulation of density functional theory\,\cite{intro14}.

Increasing interest in warm dense matter\,\cite{WDMbook,BoniRev}, an exotic state of high temperature highly compressed matter that is encountered in dense astrophysical objects (giant planet interiors, brown or white dwarfs, neutron star crusts) and ultra-fast laser heating or shock compression of metals\,\cite{intro15,intro16}, has provided the impetus for worldwide intense research activity targeted at the high density finite temperature jellium\,\cite{DornRev}. Only recently, an accurate description of the uniform electron gas has been achieved in this regime on the basis of a combination of novel path integral Monte Carlo (PIMC) methods\,\cite{intro17,intro18,FionPRL,intro19,BoniRev}.

Owing to the fact that low density finite temperature inhomogeneous electron systems are inaccessible even to contemporary state-of-the-art experiments, much less attention has been paid to the strongly coupled uniform electron liquid. In fact, there have been rather few computational and theoretical studies, in spite of the numerous intriguing physical phenomena that have been speculated to manifest themselves at strong coupling. These include the possibility of a charge-density wave instability\,\cite{intro20,intro21,intro22}, the emergence of a spin-density wave\,\cite{intro22,intro23,intro24}, the possibility of a continuous paramagnetic to ferromagnetic transition\,\cite{intro25,intro26,intro27} and the emergence of an incipient excitonic mode\,\cite{intro28,intro29}. It should be pointed out that this constitutes a particularly challenging regime for theoretical approaches due to the complex interplay between quantum effects (exchange degeneracy and diffraction), thermal excitations and strong Coulomb correlations.

The thermodynamic properties and static structure of strongly coupled electron liquids were recently probed by extensive PIMC simulations at finite temperatures\,\cite{HNCPIMC}. A dielectric formalism scheme that handles quantum mechanical effects at the random phase approximation level, assumes a frequency independent local field correction and treats strong Coulomb correlations within the classical hypernetted-chain (HNC) approximation\,\cite{HNCSTLS} was revealed to provide consistently accurate predictions for the interaction energies, although there was ample room for improvement regarding structural properties\,\cite{HNCPIMC}. From the integral equation theory of classical liquids\,\cite{IETbok1,IETbok2,IETbok3}, it is known that, as the coupling increases, the bridge diagrams, which are neglected in the HNC approximation, have an increasing importance for two-particle correlations. Hence, the HNC-based scheme is expected to gradually become inaccurate as crystallization is approached.

In a recent Letter\,\cite{BriOCP5}, we extracted the bridge functions of the classical one-component plasma at multiple thermodynamic state points, spanning the whole dense liquid region, from specially designed molecular dynamics simulations. With this input, we constructed a very accurate closed-form bridge function parametrization that covers the entire non-trivial range. This analytic description was incorporated into a novel dielectric scheme that naturally extends the HNC dielectric scheme by including the exact Coulomb bridge function. Comparison with available and new PIMC simulations of the strongly coupled electron liquid revealed that the novel scheme leads to significant improvements over the HNC-based scheme. It should be noted that this marked the first incorporation of bridge functions into a dielectric scheme, although bridge functions had been earlier embedded in variants of the so-called classical mapping method\,\cite{intro30,intro31,intro32}.

In the present work, the integral equation theory based dielectric scheme is formally introduced and its emerging self-consistent set of equations is derived. The basic assumptions and main drawbacks of this dielectric scheme are discussed together with possible further refinements. Mathematical techniques are presented that decrease the computational cost and facilitate convergence. Extensive PIMC simulations are analyzed that extend our current picture of the finite temperature strongly coupled electron liquid. A systematic comparison is carried out with other dielectric schemes and \enquote{exact} PIMC simulations in terms of interaction energies, static structure factors, static local field corrections and static density responses.

\section{Theoretical}\label{theory}

\subsection{The paramagnetic electron liquid}\label{theory:general}

\noindent The uniform electron fluid (UEF) is a homogeneous quantum mechanical system that consists of electrons that are immersed in a rigid charge neutralizing background\,\cite{IchiRep}. In other words, the UEF constitutes the quantum mechanical analogue of the classical one-component plasma (OCP). Since the electronic interactions involve not only Coulomb interactions but also exchange effects, the distinction between spin-up and spin-down electrons makes the UEF essentially a two-component system. As a consequence, the thermodynamic state points of the UEF are specified by three dimensionless parameters\,\cite{IchiRep,DornRev,BoniRev}: the \emph{quantum coupling or Brueckner parameter} $r_{\mathrm{s}}=d/a_{\mathrm{B}}$ where $d$ is the Wigner-Seitz radius $d=(4\pi{n}/3)^{-1/3}$ and $a_{\mathrm{B}}=\hbar^2/(m_{\mathrm{e}}e^2)$ the first Bohr radius, the \emph{degeneracy parameter} $\theta=k_{\mathrm{B}}T/E_{\mathrm{F}}$ where $E_{\mathrm{F}}=[(6\pi^2n^{\uparrow})^{2/3}/2](\hbar^2/m_{\mathrm{e}})$ is the Fermi energy w.r.t the Fermi wave-vector of spin-up electrons $k_{\mathrm{F}}^{\uparrow}=(6\pi^2n^{\uparrow})^{1/3}$, the \emph{spin polarization parameter} $\xi=(n^{\uparrow}-n^{\downarrow})/n$ that is constrained within $0\leq\xi\leq1$ under the spin-up choice $n^{\uparrow}\geq{n}^{\downarrow}$. In the above, $\hbar$ is the reduced Planck constant, $k_{\mathrm{B}}$ the Boltzmann constant, $e$ the electron charge, $m_{\mathrm{e}}$ the electron mass, $n=n^{\uparrow}+n^{\downarrow}$ the electron density, $T$ the temperature. In contrast to the UEF, the thermodynamic state points of the classical OCP are specified by one dimensionless parameter: the \emph{classical coupling parameter} $\Gamma=e^2/(dk_{\mathrm{B}}T)$\,\cite{IchiRep,DornRev,cOCPRev,Ott_EPJ} for which $\Gamma=2\lambda^2r_{\mathrm{s}}/\theta$ with $\lambda^3=(k_{\mathrm{F}}d)^{-3}=4/(9\pi)$.

In what follows, we shall restrict ourselves to the spin-unpolarized (paramagnetic) case, $n^{\uparrow}=n^{\downarrow}$ or $\xi=0$. The ideal (Lindhard) density response of free electrons, \emph{i.e.} in absence of Coulomb interactions, has the form\,\cite{IchiRep,DornRev,FDTbok1}
\begin{equation}
\chi_0(\boldsymbol{k},\omega)=2\int\frac{d^3q}{(2\pi)^3}\frac{f_0\left(\boldsymbol{q}\right)-f_0\left(\boldsymbol{q}+\boldsymbol{k}\right)}{\hbar\omega+\epsilon(\boldsymbol{q})-\epsilon(\boldsymbol{q}+\boldsymbol{k})+\imath\eta}\,,\label{Lindhard}
\end{equation}
with $\epsilon(\boldsymbol{q})=\hbar^2q^2/(2m_{\mathrm{e}})$ the electron kinetic energy, $\eta\to0^{+}$ owing to the adiabatic switching of the perturbation at $t\to-\infty$ that ensures causality and $f_0(\boldsymbol{q})$ the Fermi-Dirac distribution function that is given by
\begin{equation}
{f}_0\left(\boldsymbol{q}\right)=\frac{1}{\exp{\left(\displaystyle\frac{\hbar^2q^2}{2m_{\mathrm{e}}k_{\mathrm{B}}T}-\bar{\mu}\right)}+1}\,,\quad\quad\quad\quad\,\,\,\,\,\,\,\,\,\,\label{FermiDirac}
\end{equation}
with $\bar{\mu}=\beta\mu$ the reduced chemical potential ($\mu$ is the chemical potential and $1/\beta=k_{\mathrm{B}}T$) that is determined by the normalization condition $\int[d^3q/(2\pi)^3]{f}_0\left(\boldsymbol{q}\right)=n/2$.

Finally, we shall mainly focus on the high-degeneracy moderate-density range that lies beyond the warm dense matter regime\,\cite{WDMbook,DornRev,BoniRev} but simultaneously prior to the Wigner crystallization\,\cite{intro13,Wigner2}. Roughly demarcating the upper warm dense matter boundary with $r_{\mathrm{s}}\sim10$\,\cite{DornRev,BoniRev}, the above implies that we are interested in $r_{\mathrm{s}}\sim10-100$ and $\theta\sim1$. In this range, the UEF behaves as an electron liquid whose properties are determined by the interplay between strong Coulomb correlations \& quantum effects.

\subsection{The dielectric formalism}\label{theory:dielectric}

\noindent The dielectric formalism is based on (i) the quantum fluctuation dissipation theorem which connects the dynamic structure factor $S(\boldsymbol{k},\omega)$ (DSF) with the imaginary part of the exact density-density response function $\chi(\boldsymbol{k},\omega)$, (ii) the general relation that expresses $\chi(\boldsymbol{k},\omega)$ in terms of the ideal (Lindhard) density response $\chi_0(\boldsymbol{k},\omega)$ which introduces the unknown dynamic local field correction $G(\boldsymbol{k},\omega)$ (LFC), (iii) a functional relation connecting $G(\boldsymbol{k},\omega)$ with the static structure factor $S(\boldsymbol{k})$ (SSF) that is obtained by perturbative quantum / classical BBGKY approaches or non-perturbative integral equation approaches\,\cite{IchiRep,DornRev}. The first two building blocks are exact and common to all the dielectric schemes, whereas the third building block is approximate and differs.

The quantum fluctuation dissipation theorem has the general form\,\cite{FDTbok1,FDTbok2}
\begin{equation}
S(\boldsymbol{k},\omega)=-\frac{\hbar}{\pi}\frac{1}{1-e^{-\beta\hbar\omega}}\Im\{\chi(\boldsymbol{k},\omega)\}\,.\nonumber
\end{equation}
Integration over the frequency domain together with the zero frequency moment rule $S(\boldsymbol{k})=\int{S}(\boldsymbol{k},\omega)d\omega$ lead to $S(\boldsymbol{k})=-[\hbar/(2\pi{n})]\int\coth{\left[(\beta\hbar\omega)/2\right]}\Im\{\chi(\boldsymbol{k},\omega)\}d\omega$\,\cite{IchiRep,DornRev}. The extension of the $\chi(\boldsymbol{k},\omega)$ domain with the aid of analytic continuation leads to the complex valued $\widetilde{\chi}(\boldsymbol{k},z)$ and allows integral computation with contour integration techniques\,\cite{IchiMat}. Owing to the infinitely many poles of the integrand, the integral is converted to an infinite sum, \emph{i.e.}
\begin{equation}
S(\boldsymbol{k})=-\frac{1}{{n}\beta}\displaystyle\sum_{l=-\infty}^{\infty}\widetilde{\chi}(\boldsymbol{k},\imath\omega_l)\,,\quad\quad\qquad\label{Matsubaraseries}
\end{equation}
where  $\omega_l=2\pi{l}/(\beta\hbar)$ are the so-called bosonic Matsubara frequencies\,\cite{DornRev,IchiMat}.

In addition, within the linear response theory, the polarization potential approach reveals that $\chi(\boldsymbol{k},\omega)$ can always be expressed in terms of $\chi_0(\boldsymbol{k},\omega)$ and the unknown LFC as\,\cite{KuglerP,IchiRev,Ichibok}
\begin{equation}
\chi(\boldsymbol{k},\omega)=\frac{\chi_0(\boldsymbol{k},\omega)}{1-U(\boldsymbol{k})\left[1-G(\boldsymbol{k},\omega)\right]\chi_0(\boldsymbol{k},\omega)}\,,\label{densityresponseDLFC}
\end{equation}
with $U(\boldsymbol{k})=4\pi{e}^2/k^2$ the regularized Fourier transform of the Coulomb pair interaction energy.

Furthermore, all dielectric formalism schemes approximate the LFC as a SSF functional, leading to closed self-consistent approaches\,\cite{IchiRep,DornRev}. Most rigorous approaches treat quantum effects on the random phase approximation (RPA) level and correlation effects in a classical manner such as the STLS scheme\,\cite{STLSgro,STLSfin}, VS scheme\,\cite{STLSVS1,STLSVS2}, CA scheme\,\cite{STLSCA1,STLSCA2} and HNC scheme\,\cite{HNCSTLS,HNCPIMC}. These lead to a frequency independent LFC and closures of the form $G(\boldsymbol{k})\equiv{F}[S(\mathbf{k})]$. The same applies for the recently developed semi-empirical ESA scheme that utilizes established asymptotic limits and incorporates exact MC simulation results in the warm dense matter regime in order to construct an accurate frequency independent LFC\,\cite{ESAsPRL,ESAsPRB}. The qSTLS scheme constitutes a notable exception, since it captures beyond-RPA quantum effects by truncating the quantum BBGKY hierarchy within the Wigner representation at its first member with the introduction of the standard STLS closure condition\,\cite{STLSqu1,STLSqu2}. This yields a dynamic LFC and a $G(\boldsymbol{k},\omega)\equiv{F}[S(\mathbf{k}),\omega]$ closure.

Combining the above expressions, regardless of the approximation scheme, the dielectric formalism ultimately leads to a functional equation of the form
\begin{equation}
S(\boldsymbol{k})=-\displaystyle\sum_{l=-\infty}^{\infty}\frac{\widetilde{\chi}_0(\boldsymbol{k},\imath\omega_l)/(n\beta)}{1-U(\boldsymbol{k})\left\{1-{F}[S,\imath\omega_l]\right\}\widetilde{\chi}_0(\boldsymbol{k},\imath\omega_l)}\,,\label{dielectricclosurefunctional}
\end{equation}
that can be solved in an iterative manner.

\subsection{The integral equation theory of liquids}\label{theory:integral}

\noindent For one-component pair-interacting classical systems, the integral equation theory (IET) of liquids consists of two formally exact equations, the Ornstein-Zernike (OZ) integral equation and the non-linear closure equation\,\cite{IETbok1,IETbok2,IETbok3}
\begin{align}
h(r)&=c(r)+n\int c(r')h(|\boldsymbol{r}-\boldsymbol{r}'|)d^3r'\,,\label{OZequation}\\
g(r)&=\exp\left[-\beta u(r)+h(r)-c(r)+b(r)\right]\,,\label{OZclosure}
\end{align}
where $g(r)$ is the radial distribution function or pair correlation function, $h(r)=g(r)-1$ is the total correlation function, $c(r)$ is the direct correlation function and $b(r)$ is the bridge function. It is noted that capital notations $H(\boldsymbol{k}),\,C(\boldsymbol{k}),\,B(\boldsymbol{k})$ are reserved for the respective Fourier transforms. Fourier transformed radial distribution functions do not emerge in what follows and, thus, should not be confused with the frequency independent LFC $G(\boldsymbol{k})$.

The bridge function is formally defined by a virial-type expansion whose unknown coefficients are given by multi-dimensional integrals with the integrands being products of Mayer functions $f(r)=\exp{[-\beta{u}(r)]}-1$. The operation of topological reduction allows to recast the f-bond expansion as an h-bond expansion\,\cite{IETbok1,IETbok2}; a convenient resummation for long range interactions. However, slow convergence implies that rigorous computation of the exact $b[h]$ functional is impossible. Hence, most theoretical attempts have focused on the formulation of phenomenological $b[h]$ closures, typically of the form $b(h-c)$\,\cite{IETrevB}. The hypernetted-chain (HNC) approximation, which assumes that $b[h]\equiv0$, serves as a prominent example. For the classical OCP, the HNC yields near-exact structural \& thermodynamic results for $\Gamma\lesssim1$\,\cite{HNCgood}, but its accuracy strongly degrades as crystallization is approached\cite{HNCbad1,HNCbad2,HNCbad3}.

A computationally costly but far more accurate alternative concerns the indirect bridge function extraction from computer simulations\,\cite{BriCom1,BriCom2,BriCom3}. Extraction efforts for the classical OCP date back to the $80$'s\,\cite{BriOCP1,BriOCP2,BriOCP3}, but their deficiencies have been documented\,\cite{BriOCP4,BriOCP5,BriOCP6}. A very recent study managed to obtain very accurate OCP bridge functions for $17$ state points $\Gamma=10-170$ as well as construct an accurate analytic representation valid along the short range and intermediate range\,\cite{BriOCP5,BriOCP6}. In reduced units $x=r/d$, this parametrization reads as
\begin{align}
&b(x,\Gamma)=\left[1-f(x)\right]b_{\mathrm{S}}(x,\Gamma)+f(x)b_{\mathrm{I}}(x,\Gamma),\label{OCPbridgefunction}\\
&b_{\mathrm{S}}(x,\Gamma)=s_0(\Gamma)+\textstyle\sum_{i=2}^{5}s_i(\Gamma)x^{i},\nonumber\\
&b_{\mathrm{I}}(x,\Gamma)=l_0(\Gamma)\Gamma^{5/6}\exp{\left[-l_1(\Gamma)(x-1.44)-0.3x^2\right]}\times\nonumber\\&\,\,\,\,\,\,\,\,\,\,\left\{\cos{\left[l_2(\Gamma)(x-1.44)\right]}+l_3(\Gamma)\exp{\left[-3.5(x-1.44)\right]}\right\},\nonumber\\
&f(x)=0.5\left\{1+\mathrm{erf}\left[5.0\left(x-1.5\right)\right]\right\},\nonumber
\end{align}
where $b_{\mathrm{S}}(x,\Gamma)$ is the short range monotonic bridge function, $b_{\mathrm{I}}(x,\Gamma)$ is the intermediate range oscillatory decaying bridge function, $f(x)$ is a sigmoid switching function, $s_i(\Gamma)=\textstyle\sum_{j=0}^{3}s_{i}^{j}\Gamma(\ln{\Gamma})^{j},\,l_i(\Gamma)=\textstyle\sum_{j=0}^{4}l_{i}^{j}\Gamma^{1/6}(\ln{\Gamma})^{j}$ are monotonic functions of $\Gamma$ and the $s_{i}^{j},\,l_{i}^{j}$ coefficients have been tabulated\,\cite{BriOCP5,BriOCP6}. The closed system of Eqs.(\ref{OZequation},\ref{OZclosure},\ref{OCPbridgefunction}) leads to a near-exact description of strong correlations in the classical OCP.

\subsection{The IET dielectric scheme}\label{theory:functional}

\noindent The derivation of the functional closure of the IET-based scheme is based on the technique originally developed by Tanaka for the HNC-based scheme\,\cite{HNCSTLS}. The difference is that the IET non-linear closure equation for an arbitrary known $b[h]\equiv{b}(r)$ function is used instead of the HNC non-linear closure for $b(r)\equiv0$. In this section, the basic assumptions and mathematical steps will be outlined in sufficient detail for this article to remain self-contained.

The starting point is the classical ($\theta\to\infty$) fluctuation dissipation theorem, which reads as\,\cite{FDTbok2,IETbok2}
\begin{equation}
S(\boldsymbol{k},\omega)=-\frac{1}{\pi{n}\beta}\frac{\Im\{\chi(\boldsymbol{k},\omega)\}}{\omega}\,.\nonumber
\end{equation}
The integration over the frequency domain, together with the zero frequency moment rule $S(\boldsymbol{k})=\int{S}(\boldsymbol{k},\omega)d\omega$ and the $\chi(\boldsymbol{k},0)=\pi^{-1}\int[\Im\{\chi(\boldsymbol{k},\omega)\}/\omega]d\omega$ Kramers-Kronig causality relation, leads to\,\cite{ClaCor1,ClaCor2}
\begin{equation}
S(\boldsymbol{k})=-\frac{1}{{n}\beta}\chi(\boldsymbol{k},0)\,.\quad\qquad\,\,\,\,\,\nonumber
\end{equation}
Substituting for the general form of the density-density response function $\chi(\boldsymbol{k},\omega)$, approximating the LFC with a frequency independent value $G(\boldsymbol{k},\omega)\equiv{G}(\boldsymbol{k})$, employing the Maxwellian result for the static limit of the ideal classical density-density response $\chi_0(\boldsymbol{k},0)=-n\beta$, using the connecting relation $S(\boldsymbol{k})=1+nH(\boldsymbol{k})$ and invoking the Fourier transformed OZ equation $H(\boldsymbol{k})=C(\boldsymbol{k})/[1-nC(\boldsymbol{k})]$, one obtains\,\cite{ClaCor2,ClaCor3,ClaCor4}
\begin{equation}
\beta{U}(\boldsymbol{k})G(\boldsymbol{k})=C(\boldsymbol{k})+\beta{U}(\boldsymbol{k})\,.\nonumber
\end{equation}
Having established the above linear $G\equiv{F}[C]$ functional, the objective is to exploit the two exact IET equations to derive an integral $G\equiv{F}[S]$ functional. Application of the gradient operator to the non-linear closure equation and the OZ equation, see Eqs.(\ref{OZequation},\ref{OZclosure}), yields
\begin{align*}
\nabla_{\boldsymbol{r}}{h}(\boldsymbol{r})=\left[h(\boldsymbol{r})+1\right]\left[-\beta\nabla_{\boldsymbol{r}}{u}(\boldsymbol{r})+\nabla_{\boldsymbol{r}}b(\boldsymbol{r})+\right.\quad\quad\quad\,\,\,\,\,\\\left.n\int\,d^3r'\nabla_{\boldsymbol{r}}{c}(\boldsymbol{r}-\boldsymbol{r}')h(\boldsymbol{r}')\right]\,,\\
\nabla_{\boldsymbol{r}}{h}(\boldsymbol{r})=\nabla_{\boldsymbol{r}}{c}(\boldsymbol{r})+n\int\,d^3r'\nabla_{\boldsymbol{r}}{c}(\boldsymbol{r}-\boldsymbol{r}')h(\boldsymbol{r}')\,.\quad\,\,\,\,\,\,\,\,\,
\end{align*}
Equating these r.h.s., two of the convolution terms cancel out each other, resulting in
\begin{align*}
\nabla_{\boldsymbol{r}}{c}(\boldsymbol{r})+\beta\nabla_{\boldsymbol{r}}{u}(\boldsymbol{r})=-\nabla_{\boldsymbol{r}}\left[\beta{u}(\boldsymbol{r})-b(\boldsymbol{r})\right]h(\boldsymbol{r})+\quad\,\,\,\,\,\,\,\,\\\nabla_{\boldsymbol{r}}b(\boldsymbol{r})+nh(\boldsymbol{r})\int\,d^3r'\nabla_{\boldsymbol{r}}{c}(\boldsymbol{r}-\boldsymbol{r}')h(\boldsymbol{r}')\,,
\end{align*}
whose Fourier transform (after using the multiplication, convolution and differentiation properties) reads as
\begin{align*}
\boldsymbol{k}\left[{C}(\boldsymbol{k})+\beta{U}(\boldsymbol{k})\right]=\boldsymbol{k}{B}(\boldsymbol{k})+\int\,\frac{d^3q}{(2\pi)^3}\boldsymbol{q}H(\boldsymbol{k}-\boldsymbol{q})\times\,\,\,\,\,\,\,\\\left\{B(\boldsymbol{q})-\beta{U}(\boldsymbol{q})\left[1+n{H}(\boldsymbol{q})\right]+n{H}(\boldsymbol{q})\left[{C}(\boldsymbol{q})+\beta{U}(\boldsymbol{q})\right]\right\}\,.
\end{align*}
We proceed with substituting the linear $G\equiv{F}[C]$ functional relation on both sides, operating with $(\boldsymbol{k}\cdot)$ on both sides and solving for $G(\boldsymbol{k})$, which leads to
\begin{align*}
G(\boldsymbol{k})&=\frac{{B}(\boldsymbol{k})}{\beta{U}(\boldsymbol{k})}+\int\,\frac{d^3q}{(2\pi)^3}\frac{\boldsymbol{k}\cdot\boldsymbol{q}}{k^2}\frac{{U}(\boldsymbol{q})}{{U}(\boldsymbol{k})}H(\boldsymbol{k}-\boldsymbol{q})\times\,\,\,\,\,\,\,\,\\&\,\,\,\,\,\left\{\frac{B(\boldsymbol{q})}{\beta{U}(\boldsymbol{q})}-\left[1+n{H}(\boldsymbol{q})\right]+n{H}(\boldsymbol{q})G(\boldsymbol{q})\right\}\,.
\end{align*}
Using again the connecting relation $S(\boldsymbol{k})=1+nH(\boldsymbol{k})$ in order to dispose of the Fourier transformed total correlation function and substituting for $U(\boldsymbol{k})=4\pi{e}^2/k^2$, ultimately results in the sought-for IET functional
\begin{align}
G_{\mathrm{IET}}(\boldsymbol{k})&=\frac{{B}(\boldsymbol{k})}{\beta{U}(\boldsymbol{k})}-\frac{1}{n}\int\,\frac{d^3q}{(2\pi)^3}\frac{\boldsymbol{k}\cdot\boldsymbol{q}}{q^2}[S(\boldsymbol{k}-\boldsymbol{q})-1]\times\nonumber\\&\,\,\,\,\,\left\{1-\frac{B(\boldsymbol{q})}{\beta{U}(\boldsymbol{q})}-\left[G(\boldsymbol{q})-1\right]\left[S(\boldsymbol{q})-1\right]\right\}\,.\label{IETclosurefunctional}
\end{align}

As expected, the IET functional collapses to the HNC functional for $b(r)\equiv0$, which explicitly reads as\,\cite{HNCSTLS,HNCPIMC}
\begin{align}
G_{\mathrm{HNC}}(\boldsymbol{k})&=-\frac{1}{n}\int\,\frac{d^3q}{(2\pi)^3}\frac{\boldsymbol{k}\cdot\boldsymbol{q}}{q^2}[S(\boldsymbol{k}-\boldsymbol{q})-1]\times\quad\qquad\nonumber\\&\,\,\,\,\,\left\{1-\left[G(\boldsymbol{q})-1\right]\left[S(\boldsymbol{q})-1\right]\right\}\,.\label{HNCclosurefunctional}
\end{align}
It is rather straightforward to express the IET functional in terms of the HNC functional,
\begin{align*}
G_{\mathrm{IET}}(\boldsymbol{k})&=\frac{1}{n}\int\,\frac{d^3q}{(2\pi)^3}\frac{\boldsymbol{k}\cdot\boldsymbol{q}}{q^2}\frac{B(\boldsymbol{q})}{\beta{U}(\boldsymbol{q})}[S(\boldsymbol{k}-\boldsymbol{q})-1]+\,\,\,\,\,\nonumber\\&\,\,\,\,\,\frac{{B}(\boldsymbol{k})}{\beta{U}(\boldsymbol{k})}+G_{\mathrm{HNC}}(\boldsymbol{k})\,.
\end{align*}
Considering the short range of the OCP bridge function and the long range of Coulomb interactions, this relation suggests that the IET scheme results in small corrections to the LFC of the HNC scheme at the long and intermediate wavelength regions.

\section{Numerical}\label{numerical}

\subsection{On the treatment of the IET closure functional}\label{numerical:bipolar}

\noindent Considering the isotropy of homogeneous electron liquids, the IET closure functional $G_{\mathrm{IET}}(\boldsymbol{k})$ is formally expressed as a triple integral of the type $\int{d}^3q(\boldsymbol{k}\cdot\boldsymbol{q})I(|\boldsymbol{k}-\boldsymbol{q}|)J(|\boldsymbol{q}|)$, see Eq.(\ref{IETclosurefunctional}). Such a triple integral type is also encountered in the course of the Wertheim-Thiele derivation of the exact solution of the Percus-Yevick approximation for hard spheres\,\cite{bipola1,bipola2}, the Laplace transform based derivation of the exact solution of the soft mean spherical approximation for classical plasmas\,\cite{bipola3,bipola4} and the mathematical treatment of the asymptotic memory kernel in mode coupling theories of classical supercooled liquids\,\cite{bipola5,bipola6,bipola7}. This triple integral can be directly converted to a double integral with the introduction of azimuthally expanded two-center bipolar coordinates, which is equivalent to the sequential transformations $\boldsymbol{q}=\boldsymbol{p}+(1/2)\boldsymbol{k}$, spherical coordinates for $\boldsymbol{p}$ assuming $\boldsymbol{k}||\hat{\boldsymbol{z}}$ without loss of generality, and $u=|\boldsymbol{p}+(1/2)\boldsymbol{k}|\,,w=|\boldsymbol{p}-(1/2)\boldsymbol{k}|$ with a surface element identity $p^2\sin{\theta}d\theta{d}p=(uw/k)dudw$. Ultimately, the IET closure functional becomes
\begin{align*}
G&(k)=\frac{{B}(k)}{\beta{U}(k)}-\frac{1}{n}\frac{1}{8\pi^2k}\int_0^{\infty}\left\{-\frac{B(u)}{\beta{U}(u)}+1-\left[{G}(u)-1\right]\right.\\&\left.\times\left[S(u)-1\right]\right\}udu\int_{|u-k|}^{u+k}\frac{u^2-w^2+k^2}{u^2}w\left[S(w)-1\right]dw\,.
\end{align*}
Utilizing dimensionless variables for all the wave-vectors, \emph{i.e.} $y\to{u}/k_{\mathrm{F}}$, $z\to{w}/k_{\mathrm{F}}$, $x\to{k}/k_{\mathrm{F}}$ and rearranging the integrands, the double integral expression for the IET closure functional reads as
\begin{align*}
G&(x)=\frac{{B}(x)}{\beta{U}(x)}+\frac{3}{8x}\int_0^{\infty}\left\{-\frac{B(y)}{\beta{U}(y)}+1-\left[{G}(y)-1\right]\times\right.\\&\left.\left[S(y)-1\right]\right\}\frac{dy}{y}\int_{|y-x|}^{y+x}\left(z^2-y^2-x^2\right)z\left[S(z)-1\right]dz\,.
\end{align*}
To our knowledge, bipolar coordinates have never been utilized in earlier applications of the CA scheme\,\cite{STLSCA1,STLSCA2} and the HNC scheme\,\cite{HNCSTLS,HNCPIMC}, whose LFC closure functional is also formally expressed as a triple integral of the type $\int{d}^3q(\boldsymbol{k}\cdot\boldsymbol{q})I(|\boldsymbol{k}-\boldsymbol{q}|)J(|\boldsymbol{q}|)$, although this would lead to a drastic reduction of the computational cost without invoking extra approximations. This comes in stark contrast to the so-called modified versions of the CA scheme and the HNC scheme\,\cite{HNCSTLS,STLSMCA}, where the triple integral is converted to a single integral after replacing $S(\boldsymbol{k}-\boldsymbol{q})$ by an ad hoc Yukawa screening function of a characteristic screening wave-number that is determined by an interaction energy constraint.

\subsection{On the convergence of the infinite Matsubara series}\label{numerical:series}

\noindent Independent of the dielectric scheme (STLS, CA, qSTLS, HNC, IET), the Matsubara summation of Eq.(\ref{dielectricclosurefunctional}) is slowly converging, especially for small values of the degeneracy parameter $\theta$. The implementation of mathematical tricks that speed-up the convergence rate is especially important for computationally heavy schemes such as the HNC, IET and qSTLS.

A significant speed-up of the rate of convergence can be achieved by splitting the Hartree-Fock SSF, \emph{i.e.} the SSF in absence of Coulomb pair interactions $U(k)=0$, since its respective Matsubara summation can be calculated exactly\,\cite{IchiMat,STLSfin,STLSqu2,STLSqu3}. Introducing the auxiliary complex function $\Phi(k,z)=-(2E_{\mathrm{F}})/(3n)\widetilde{\chi}_0(k,z)$ and normalized wave-vectors $x\to{k}/k_{\mathrm{F}}$, Eq.(\ref{dielectricclosurefunctional}) becomes
\begin{align*}
&S(x)=S_{\mathrm{HF}}(x)-\displaystyle\sum_{l=-\infty}^{\infty}\frac{\displaystyle\frac{6}{\pi}\lambda{r}_{\mathrm{s}}\theta\frac{1}{x^2}\left[1-G(x)\right]\Phi^2(x,l)}{1+\displaystyle\frac{4}{\pi}\lambda{r}_{\mathrm{s}}\frac{1}{x^2}\left[1-G(x)\right]\Phi(x,l)}\,,
\end{align*}
where, courtesy of logarithm properties and the product $\prod_{n=-\infty}^{+\infty}\left(a^2+\pi^2n^2\right)/\left(b^2+\pi^2n^2\right)=\sinh^2{(a)}/\sinh^2{(b)}$, the Hartree-Fock SSF $S_{\mathrm{HF}}(x)=(3/2)\theta\sum_{l=-\infty}^{\infty}\Phi(x,l)$ is given by the integral\,\cite{IchiMat,STLSfin,STLSqu3}
\begin{align*}
&S_{\mathrm{HF}}(x)=1-\frac{3\theta}{4x}\int_0^{\infty}ydy\frac{\displaystyle\ln{\left\{\frac{1+\exp{\left[\bar{\mu}-\frac{(y-x)^2}{\theta}\right]}}{1+\exp{\left[\bar{\mu}-\frac{(y+x)^2}{\theta}\right]}}\right\}}}{\exp{\left(\frac{y^2}{\theta}-\bar{\mu}\right)}+1}\,.
\end{align*}

Convergence can be further accelerated by splitting the square of the high frequency - short wavelength asymptotic form $(l,q\to\infty)$ of the auxiliary complex function $\Phi_{\infty}(x,l)=(4/3)x^2/[x^4+(2\pi{l}\theta)^2]+\mathcal{O}(x^{-4},l^{-4})$, since its respective Matsubara summation can also be calculated exactly\,\cite{IchiMat,STLSfin,STLSqu2}. This leads to
\begin{align*}
&S(x)=S_{\mathrm{HF}}(x)-S_{\infty}(x)-\frac{6}{\pi}\lambda{r}_{\mathrm{s}}\theta\frac{1}{x^2}\left[1-G(x)\right]\times\\&\displaystyle\sum_{l=-\infty}^{\infty}\left\{\frac{\Phi^2(x,l)}{1+\displaystyle\frac{4}{\pi}\lambda{r}_{\mathrm{s}}\frac{1}{x^2}\left[1-G(x)\right]\Phi(x,l)}-\Phi_{\infty}^2(x,l)\right\}\,.
\end{align*}
where, by differentiating both sides of the known formula $\sum_{n=-\infty}^{+\infty}(x^4+a^2n^2)^{-1}=[\pi/(ax^2)]\coth{\left(\pi{x}^2/a\right)}$ with respect to $x$, the residual SSF correction that is defined by $S_{\infty}(x)=(6/\pi)\lambda{r}_{\mathrm{s}}\theta(1/{x}^{2})\left[1-G(x)\right]\sum_{l=-\infty}^{\infty}\Phi_{\infty}^2(x,l)$ is given by\,\cite{IchiMat,STLSfin}
\begin{align*}
&S_{\infty}(x)=\frac{4}{3\pi}\frac{\lambda{r}_{\mathrm{s}}}{\theta}\frac{1-G(x)}{x^2}\left[\mathrm{csch}^2{\left(\frac{x^2}{2\theta}\right)}+\frac{2\theta}{x^2}\coth{\left(\frac{x^2}{2\theta}\right)}\right]\,,
\end{align*}
where $\coth{(\cdot)}$ is the hyperbolic cotangent and $\mathrm{csch}{(\cdot)}$ the hyperbolic cosecant.

On the practical side, with an accuracy goal of $0.001\%$ in the interaction energies, the above tricks speed-up the convergence rate by more than two orders of magnitude. To be more specific, for $r_{\mathrm{s}}=100$ and for $\theta=0.5$ or $4$, this accuracy goal is well achieved within $l=512$, when these two splitting procedures are implemented. On the other hand, for $r_{\mathrm{s}}=100$, the interaction energy accuracy at $l=51200$ is merely $0.1\%$ for $\theta=4$ and $0.7\%$ for $\theta=0.5$, when no splitting is implemented.

\subsection{On the asymptotic convergence of the IET local field correction}\label{numerical:asymptotic}

\noindent From the general SSF decomposition in terms of $S_{\mathrm{HF}}(x)$ and $S_{\infty}(x)$, it is rather straightforward to prove that the frequency independent LFC assumption $G(\boldsymbol{k},\omega)\equiv{G}(\boldsymbol{k})$ implies the  $(3\pi/8)x^4[1-S(x)]=\lambda{r}_{\mathrm{s}}[1-G(x)]$ asymptotic limit\,\cite{IchiMat}. When combined with Kimball's expression\,\cite{asympt1} $\partial{g}(\widetilde{r}=0)/\partial\widetilde{r}=(3\pi/8)\lim_{x\to\infty}x^4[1-S(x)]$ with $\widetilde{r}=rk_{\mathrm{F}}$ and the cusp relation\,\cite{asympt1,asympt2} $\partial{g}(\widetilde{r}=0)/\partial\widetilde{r}=\lambda{r}_{\mathrm{s}}g(0)$, this yields the general asymptotic condition $G(x\to\infty)=1-g(0)$\,\cite{IchiMat,asympt3} where $g(0)$ is the contact or on-top value of the radial distribution function. This asymptotic condition is often considered as a self-consistency condition, since it solely originates from the first two building blocks of the dielectric formalism and needs to be independently satisfied by the third building block (closure functional). This has been shown to be valid in the case of the STLS scheme\,\cite{IchiMat} and can also be confirmed for other dielectric schemes (VS, CA, HNC) as well as for the IET scheme, given the rapid ${B}(x)/\beta{U}(x)$ decay to zero. Note that, in the strongly coupled electron liquid regime, $g(0)\simeq0$ is expected\,\cite{ESAsPRL,asympt4} leading to $G(x\to\infty)\simeq1$.

Nevertheless, the implicit form of the IET LFC can strongly inhibit its proper numerical convergence at short wavelengths. In particular, the converged LFC solution has been consistently observed to exhibit a short wavelength dependence which abruptly drops from near-unity to near-zero. The unphysical asymptotic behavior always takes place close to the wavenumber cut-off considered in the integrations, regardless of its actual value. It translates to interaction energy errors of the order of $0.005\%$, which exceed our $0.001\%$ accuracy goal.

In order to achieve proper convergence of the IET local field correction in the short wavelength limit, it is highly beneficial to split the STLS LFC from the IET LFC. The STLS LFC emerges by setting $B(k)\equiv0$ and $G(k)\equiv0$ in the r.h.s. of the IET LFC. The standard single integral form of the STLS LFC is recovered from the double integral form of the IET LFC by employing the change of variables $(y,z)\to(t,s)$ of the form $y=\sqrt{x^2+s^2-2xst}$, $z=s$ and carrying out the $t-$ integration. This leads to
\begin{align*}
G&(x)=G_{1}(x)+G_{2}(x)\,,\\
G&_{1}(x)=-\frac{3}{4}\int_{0}^{\infty}s^2\left[S(s)-1\right]\left[1+\frac{x^2-s^2}{2xs}\ln{\left|\frac{x+s}{x-s}\right|}\right]ds\,,\\
G&_{2}(x)=\frac{{B}(x)}{\beta{U}(x)}-\frac{3}{8x}\int_0^{\infty}\left\{\frac{B(y)}{\beta{U}(y)}+\left[{G}(y)-1\right]\times\right.\\&\quad\quad\,\,\,\,\left.\left[S(y)-1\right]\right\}\frac{dy}{y}\int_{|y-x|}^{y+x}\left(z^2-y^2-x^2\right)z\left[S(z)-1\right]dz\,.
\end{align*}
In the above, $G_1(x)$ denotes the STLS LFC with a numerical asymptotic behavior $G_1(x\to\infty)\simeq1$ and $G_2(x)$ denotes the residual IET LFC with a numerical asymptotic behavior $G_2(x\to\infty)=0$. Thus, the IET LFC short wavelength limit now correctly converges to a value close to unity.

\subsection{Structure and details of the IET algorithm}\label{numerical:algorithm}

\begin{figure}
	\centering
	\includegraphics[width=3.15in]{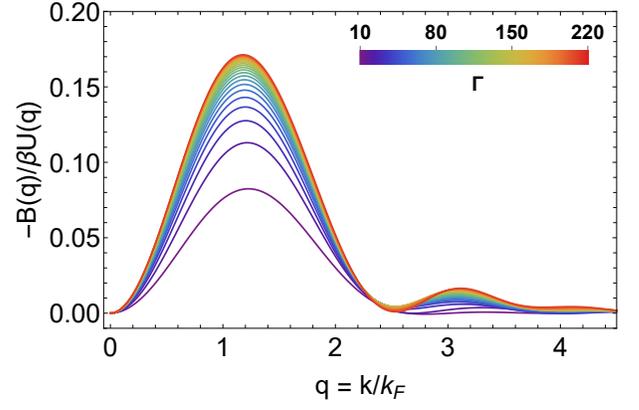}
	\caption{The non-trivial $-B(q)/\beta{U}(q)$ contribution in the normalized wavenumber range $q=k/{k}_{\mathrm{F}}\leq4.5$ for different values of the classical coupling parameter $\Gamma=2\lambda^2(r_{\mathrm{s}}/\theta)$. Results for $\Gamma=10-220$. It should be noted that the classical OCP bridge function extrapolates smoothly in the range $\Gamma=170-220$. For the computation, Eq.(\ref{IETSTLSfinal4}) was combined with Eq.(\ref{OCPbridgefunction}). The numerical integration was performed with the doubly adaptive Clenshaw-Curtis quadrature method.}\label{fig:bridgefunctions}
\end{figure}

\noindent The closed normalized set of equations for the IET-based dielectric scheme comprises of: the normalization condition of the Fermi-Dirac energy distribution function [see Eq.(\ref{IETSTLSfinal1}), Sec.\ref{theory:general}], the ideal Lindhard density response expressed through the auxiliary complex function $\Phi(x,l)$ and evaluated at the imaginary Matsubara frequencies $\omega_l$ including the static limit [see Eqs.(\ref{IETSTLSfinal2},\ref{IETSTLSfinal3}), Sec.\ref{theory:general}], the Fourier transform of the classical OCP bridge function [see Eq.(\ref{IETSTLSfinal4}), Sec.\ref{theory:integral}], the infinite Matsubara summation expression for the SSF after separating the Hartree-Fock SSF and also the asymptotic component [see Eq.(\ref{IETSTLSfinal5}), Sec.\ref{numerical:series}], the IET LFC double integral expression after utilizing bipolar coordinates and splitting the STLS LFC single integral expression [see Eq.(\ref{IETSTLSfinal6}), Secs.\ref{numerical:bipolar},\ref{numerical:asymptotic}].
\begin{align}
&\int_0^{\infty}\frac{\sqrt{z}dz}{\exp{\left(z-\bar{\mu}\right)}+1}=\frac{2}{3}\theta^{-3/2}\,,\label{IETSTLSfinal1}\\
&\Phi(x,l)=\frac{1}{2x}\int_0^{\infty}\frac{y}{\exp{\left(\frac{y^2}{\theta}-\bar{\mu}\right)}+1}\times\label{IETSTLSfinal2}\\&\quad\quad\qquad\ln{\left[\frac{\left(x^2+2xy\right)^2+\left(2\pi{l}{\theta}\right)^2}{\left(x^2-2xy\right)^2+\left(2\pi{l}\theta\right)^2}\right]}dy\,,\nonumber\\
&\Phi(x,0)=\frac{1}{\theta{x}}\int_0^{\infty}\frac{y\exp{\left(\frac{y^2}{\theta}-\bar{\mu}\right)}}{\left[\exp{\left(\frac{y^2}{\theta}-\bar{\mu}\right)}+1\right]^2}\times\label{IETSTLSfinal3}\\&\quad\quad\qquad\left[\left(y^2-\frac{x^2}{4}\right)\ln{\left|\frac{2y+x}{2y-x}\right|}+xy\right]dy\,,\nonumber\\
&\frac{B(q)}{\beta{U}(q)}=\frac{9\pi}{8}\frac{\theta}{r_{\mathrm{s}}}q\int_0^{\infty}yb\left(y,2\lambda^2\frac{r_{\mathrm{s}}}{\theta}\right)\sin{\left(\frac{q}{\lambda}y\right)}dy\,,\label{IETSTLSfinal4}\\
&S(x)=S_{\mathrm{HF}}(x)-S_{\infty}(x)-\frac{6}{\pi}\lambda{r}_{\mathrm{s}}\theta\frac{1}{x^2}\left[1-G(x)\right]\times\label{IETSTLSfinal5}\\&\quad\quad\quad\displaystyle\sum_{l=-\infty}^{\infty}\left\{\frac{\Phi^2(x,l)}{1+\displaystyle\frac{4}{\pi}\lambda{r}_{\mathrm{s}}\frac{1}{x^2}\left[1-G(x)\right]\Phi(x,l)}-\Phi_{\infty}^2(x,l)\right\}\,,\nonumber\\
&G(x)=-\frac{3}{4}\int_{0}^{\infty}s^2\left[S(s)-1\right]\left[1+\frac{x^2-s^2}{2xs}\ln{\left|\frac{x+s}{x-s}\right|}\right]ds+\nonumber\\&\quad\quad\quad\,\,\,\frac{{B}(x)}{\beta{U}(x)}-\frac{3}{8x}\int_0^{\infty}\left\{\frac{B(y)}{\beta{U}(y)}+\left[{G}(y)-1\right]\times\right.\label{IETSTLSfinal6}\\&\quad\quad\quad\,\,\left.\left[S(y)-1\right]\right\}\frac{dy}{y}\int_{|y-x|}^{y+x}\left(z^2-y^2-x^2\right)z\left[S(z)-1\right]dz\,.\nonumber
\end{align}
Concerning the origin of Eq.(\ref{IETSTLSfinal4}), within the classical $x=r/d$ or $q=kd$ normalization, it is rather straightforward to show that $B(q)/\beta{U}(q)=(q/\Gamma)\int_0^{\infty}xb(x,\Gamma)\sin{(qx)}dx$ for the ratio of spatial Fourier transforms, where $b(x,\Gamma)$ is directly adopted from Eq.(\ref{OCPbridgefunction}). Naturally, the above expression needs to be translated to the quantum $x=rk_{\mathrm{F}}$ or $q=k/k_{\mathrm{F}}$ normalization, which is formally equivalent to the substitution $q\to{q}/\lambda$. Finally, it is apparent that the utilization of classical OCP bridge functions necessitates a mapping of the quantum states ($r_{\mathrm{s}},\theta$) to classical states ($\Gamma$) via $\Gamma=2\lambda^2(r_{\mathrm{s}}/\theta)$. The $B(q)/\beta{U}(q)$ contribution has been illustrated in Fig.\ref{fig:bridgefunctions}.

The accuracy goal was set to $0.001\%$ with respect to the interaction energies. All improper integrals were numerically evaluated with the doubly adaptive Clenshaw-Curtis quadrature method, as implemented in the GNU Scientific Library, with a $0.1k_{\mathrm{F}}$ grid resolution and with a $40k_{\mathrm{F}}$ upper cut-off. The only exception was the complete Fermi-Dirac integral $I_{1/2}(\bar{\mu})$, which is implemented as a special function in the GNU Scientific Library. It should be noted that non-adaptive quadrature rules were confirmed to require much denser grid spacing $<0.005k_{\mathrm{F}}$ in order to satisfy the accuracy goal. The infinite Matsubara summation was truncated at $|l|=512$. Convergence studies were carried out, for representative quantum coupling parameters and degeneracy parameters, in order to ensure that the chosen resolution and cutoffs do not affect the thermodynamic and structural results.

The iteration cycle proceeds as follows:\,\textbf{(i)} The reduced chemical potential $\bar{\mu}$ is calculated from Eq.(\ref{IETSTLSfinal1}) using a bisection root-finding algorithm. \textbf{(ii)} The Lindhard density responses ($0\leq{l}\leq512$) are computed from Eqs.(\ref{IETSTLSfinal2},\ref{IETSTLSfinal3}). \textbf{(iii)} The Fourier transform of the classical OCP bridge function is computed from Eq.(\ref{IETSTLSfinal4}) and stored. \textbf{(iv)} With the RPA's LFC as an initial guess, the SSF is evaluated from Eq.(\ref{IETSTLSfinal5}). \textbf{(v)} These LFC and SSF are substituted in the r.h.s. of Eq.(\ref{IETSTLSfinal6}) for an initial evaluation of the IET LFC. \textbf{(vi)} The last two steps are repeated, until the absolute relative difference between two successive IET LFC evaluations is smaller than $10^{-5}$ for all the grid points. Starting from the RPA solution, convergence is typically reached within $200$ iterations. Especially for state points with $r_{\mathrm{s}}>100$ and $\theta<1.0$, Broyles' technique of mixing iterates\,\cite{HNCbad1,mixingB} was necessary to speed up and sometimes even to achieve convergence.

\section{Computational}\label{computational}

\noindent To compute accurate benchmark data for our new IET scheme, we have carried out direct PIMC simulations\,\cite{PIMCR01,PIMCR02,PIMCR03} of the UEF without any nodal restrictions\,\cite{PIMCR04}. As a consequence, our simulations are afflicted with the notorious fermion sign problem\,\cite{PIMCR05}, which, in general, leads to an exponential increase in the compute time with increasing the system size $N$ or decreasing the temperature $T$; the reader is addressed to Refs.\cite{PIMCR06,PIMCR07} for topical and accessible review articles. In practice, however, the sign problem is not severe as quantum exchange effects are effectively reduced by the strong Coulomb repulsion in the strongly coupled electron liquid regime.

The basic idea behind the PIMC method is to evaluate the (canonical, i.e. system-size $N$, volume $V$ and inverse temperature $\beta=1/k_\text{B}T$ are fixed) partition function in coordinate space, which gives
\begin{eqnarray}\label{eq:Z}
Z_{\beta,N,V} &=& \frac{1}{N^\uparrow! N^\downarrow!} \sum_{\sigma^\uparrow\in S_{N^\uparrow}} \sum_{\sigma^\downarrow\in S_{N^\downarrow}} \textnormal{sgn}(\sigma^\uparrow,\sigma^\downarrow)\\\nonumber & & \times \int d\mathbf{R} \bra{\mathbf{R}} e^{-\beta\hat H} \ket{\hat{\pi}_{\sigma^\uparrow}\hat{\pi}_{\sigma^\downarrow}\mathbf{R}}\,.
\end{eqnarray}
Specifically, $\mathbf{R}=(\mathbf{r}_1,\dots,\mathbf{r}_{N})^T$ contains the coordinates of all $N=N^\uparrow + N^\downarrow$ electrons, and, due to the antisymmetry of the fermionic density matrix under the exchange of particle coordinates, we have to explicitly take the sums over all elements $\sigma_i$ of the respective permutation group $S_{N^i}$, $i\in\{\uparrow,\downarrow\}$, where the sign function $\textnormal{sgn}(\sigma^\uparrow,\sigma^\downarrow)$ gives positive (negative) unity for an even (odd) number of pair permutations.
Further, the operators $\hat{\pi}_{\sigma^i}$ realize the particular permutations for a corresponding element $\sigma^i$.

The problem with Eq.(\ref{eq:Z}) is that the matrix elements of the density operator $e^{-\beta\hat H}$ cannot be readily evaluated, since the kinetic ($\hat K$) and potential ($\hat V$) contributions to the Hamiltonian $\hat H = \hat K + \hat V$ do not commute, $e^{-\beta\hat H}\neq e^{-\beta\hat K}e^{-\beta\hat V}$. In order to overcome this obstacle, we utilize the exact semi-group property of $\hat\rho$
\begin{eqnarray}\label{eq:group}
e^{-\beta\hat H} = \prod_{\alpha=0}^{P-1} e^{-\epsilon\hat H}\,,
\end{eqnarray}
where the definition $\epsilon=\beta/P$ has been employed. Applying Eq.(\ref{eq:group}) to Eq.(\ref{eq:Z}) and inserting $P-1$ unity operators of the form $\hat{1}=\int\textnormal{d}\mathbf{R}_\alpha \ket{\mathbf{R}_\alpha}\bra{\mathbf{R}_\alpha}$ leads to the intermediate result
\begin{eqnarray}\label{eq:Z_modified}
Z_{\beta,N,V}&=&\frac{1}{N^\uparrow! N^\downarrow!} \sum_{\sigma^\uparrow\in S_{N^\uparrow}} \sum_{\sigma^\downarrow\in S_{N^\downarrow}} \textnormal{sgn}(\sigma^\uparrow,\sigma^\downarrow)\\\nonumber & & \times \int d\mathbf{R}_0\dots d\mathbf{R}_{P-1}
\bra{\mathbf{R}_0}e^{-\epsilon\hat H}\ket{\mathbf{R}_0}\\\nonumber & & \times \bra{\mathbf{R}_1}e^{-\epsilon\hat H}\ket{\mathbf{R}_1} \dots 
\bra{\mathbf{R}_{P-1}} e^{-\epsilon\hat H} \ket{\hat{\pi}_{\sigma^\uparrow}\hat{\pi}_{\sigma^\downarrow}\mathbf{R}_0}\,,
\end{eqnarray}
which is still exact. Evidently, Eq.(\ref{eq:Z_modified}) requires the evaluation of $P$ density matrices, but at $P$ times the temperature. For a sufficiently large $P$, each of these factors can be straightforwardly evaluated using a suitable high-temperature approximation, like the primitive factorization $e^{-\epsilon\hat H}\approx e^{-\epsilon\hat K}e^{-\epsilon\hat V}$. In fact, the factorization error of the latter decays as $\sim 1/P^2$\,\cite{Kleiner}, and the convergence in the limit of $P\to\infty$ is ensured by the celebrated Trotter formula\,\cite{Trotter}
\begin{eqnarray}\label{eq:trotter}
\lim_{P\to\infty} \left( e^{-\epsilon\hat{K}} e^{-\epsilon\hat{V}} \right)^P = e^{-\beta(\hat K + \hat V)}\ .
\end{eqnarray}
In practice, we find $P=200$ sufficient to reduce the factorization error substantially below the noise level of the Monte Carlo simulation. For completeness, we note that Eq.(\ref{eq:trotter}) only holds for the case of potentials $\hat V$ that are bounded from below, which is indeed the case for the UEF. Attractive potentials like the Coulomb interaction between positive and negative charges require a modified procedure such as the pair approximation, which is discussed in detail for instance in Ref.\cite{Militze}. In addition, we note that higher-order factorizations of the thermal density matrix that converge as $\sim1/P^4$\,\cite{Ksakkos,cpp2019} and even $\sim1/P^{6-8}$\,\cite{zillich} have been presented, although we do not find them necessary for the present conditions.

The final result for the PIMC partition function can then be written in abbreviated form as
\begin{eqnarray}\label{eq:Z_final}
Z = \int\textnormal{d}\mathbf{X}\ W(\mathbf{X})\,,
\end{eqnarray}
where the integration over the $\mathbf{X}=(\mathbf{R}_0,\dots,\mathbf{R}_{P-1})^T$ meta-variable also contains the summation over all the possible permutations. Furthermore, the weight function $W(\mathbf{X})$ can be readily evaluated for each individual configuration $\mathbf{X}$ and contains contributions from the potential energy and from the free thermal density matrix; see for instance Ref.\cite{PIMCR01} for an accessible review article. Evidently, Eq.(\ref{eq:Z_final}) requires the evaluation of a $d=3PN$-dimensional integral, which easily leads to $d\sim10^{4}$ in the present study. While this renders the application of standard quadrature methods impractical due to the well-known curse of dimensionality, the Metropolis Monte Carlo method\,\cite{metropo} does not suffer from this drawback.

\begin{figure}
\centering
\includegraphics[width=3.40in]{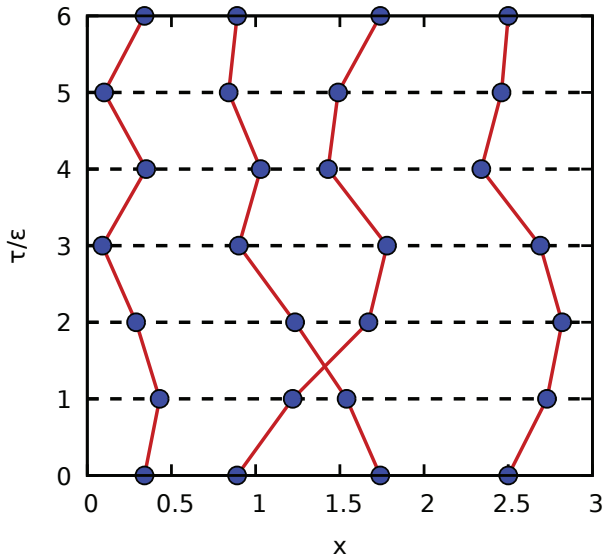}
\caption{\label{fig:PIMC} Illustration of the PIMC method. A configuration of $N=4$ particles is shown in the $\tau$-$x$-plane for $P=6$ high-temperature factors. Notice the single pair exchange of the two particles in the center leading to a negative sign for this particular configuration, $W(\mathbf{X})<0$.}
\end{figure}

A graphical illustration of the PIMC approach is shown in Fig.\ref{fig:PIMC}. Specifically, we show a configuration $\mathbf{X}$ of $N=4$ electrons with $P=6$ high-temperature factors. It is evident that each particle is represented by a closed (i.e., $\beta$-periodic) path of particle coordinates in the imaginary time $\tau\in[0,\beta]$ (strictly speaking, it is $\tau\in\hbar/i[0,\beta]$, but it is conventional to drop the pre-factor for simplicity). This, in turn, corresponds to the famous \emph{classical isomorphism}\,\cite{chandle}, where the complicated quantum many-body system of interest is mapped onto an effective classical system of interacting ring-polymers.

An additional difficulty arises due to the indistinguishable nature of electrons of the same spin-species, which requires us to sample all the possible permutations of particle coordinates. Within the PIMC picture, the latter manifest as so-called permutation cycles\,\cite{permcyc}, which are trajectories comprising more than a single particle. An example for such a permutation cycle can be identified at the center of Fig.\ref{fig:PIMC}. As a consequence of the depicted pair exchange, we have to move twice through the imaginary time to return to the point of origin. This, in turn, results in a negative sign of the corresponding configuration weight, $W(\mathbf{X})<0$, and thereby contributes to the aforementioned fermion sign problem.

In practice, we construct a Markov chain of random configurations $\{\mathbf{X}_i\}$ using a canonical adaption\,\cite{mezzaca} of the worm algorithm by Boninsegni \emph{et al.}\,\cite{bonins1,bonins2}.

\begin{figure}
\centering
\includegraphics[width=3.40in]{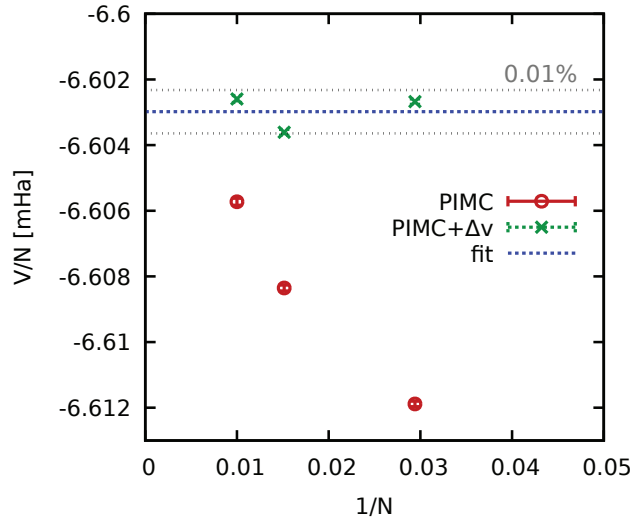}
\caption{\label{fig:v_FSC} The finite-size extrapolation of the PIMC data. The UEF interaction energy per particle is shown at $r_{\mathrm{s}}=125$ and $\theta=2$ as a function of the inverse electron number. The red circles depict the raw PIMC results for different numbers of electrons $N$ and the green crosses have been obtained by adding onto the former the finite-size correction from Eq.(\ref{eq:FSC_v}). The horizontal blotted blue line shows the average value of the corrected data points, and the light dotted grey lines indicate a deviation of $0.01\%$.}
\end{figure}

Let us next consider the PIMC estimation of the interaction energy per particle $V_N/N$, which is shown in Fig.\ref{fig:v_FSC} for $r_{\mathrm{s}}=125$ and $\theta=2$. In particular, PIMC simulations are, by default, only possible for a finite number of electrons $N$. The raw PIMC data for three different system sizes are shown as the red circles in Fig.\ref{fig:v_FSC} and exhibit a significant dependence on $N$. In practice, however, we are interested in the thermodynamic limit, \emph{i.e.}, in the limit of an infinite number of particles with the density being constant,
\begin{eqnarray}\label{eq:TDL}
v = \lim_{N\to\infty} \left. \frac{V_N}{N}\right|_{r_{\mathrm{s}},\theta}\,.
\end{eqnarray}
To eliminate the difference between the PIMC data for $V_N/N$ and Eq.(\ref{eq:TDL}), we use the finite-size correction by Chiesa \emph{et al.}\,\cite{ChiesaL}, which has subsequently been adapted to finite temperatures in Ref.\cite{parame4},
\begin{eqnarray}\label{eq:FSC_v}
\Delta v(N)=\frac{\omega_{\mathrm{p}}}{4N}\textnormal{coth}\left(\frac{\beta\omega_{\mathrm{p}}}{2}\right)\,,
\end{eqnarray}
where the plasma frequency is given by $\omega_{\mathrm{p}}=\sqrt{3/r_{\mathrm{s}}^3}$ in Hartree atomic units. To be more specific, Eq.(\ref{eq:FSC_v}) is based on the insight that the system-size dependence of $V_N/N$ is mainly the consequence of the approximation of a continuous integral, see Eq.(\ref{interactionenergies}) below, by the sum over reciprocal lattice vectors due to the momentum quantization in a finite simulation cell. To the first order, this discretization error can be approximated by utilizing the exact long wave-length limit of the static structure factor\,\cite{kuglerb}
\begin{eqnarray}
\lim_{k\to0}S(\mathbf{k}) = \frac{\mathbf{k}^2}{2\omega_{\mathrm{p}}}\textnormal{coth}\left(\frac{\beta\omega_{\mathrm{p}}}{2}\right)\,;
\end{eqnarray}
a detailed derivation is beyond the scope of the present work and has been presented by Drummond \emph{et al.}\,\cite{Drummon}. Adding the finite-size correction given in Eq.(\ref{eq:FSC_v}) to the PIMC data leads to the green crosses in Fig.\ref{fig:v_FSC}. Evidently, the bulk of the finite-size errors have been removed, and the corrected data points fall into an interval of $0.01\%$ (horizontal light grey lines) around their common average value (horizontal blue line). All the PIMC results for the interaction energy that are shown in this work have been obtained by following this procedure.

For completeness, it should be mentioned that finite-size effects are substantially more pronounced in the warm dense matter regime and that the simple first-order correction from Eq.(\ref{eq:FSC_v}) breaks down\,\cite{parame6,DornRev}. A recent investigation of finite-size effects of a uniform electron gas at extreme densities and temperatures has been presented by Dornheim and Vorberger\,\cite{DornVor}.

\section{Results}\label{results}

\noindent Here, we compare the paramagnetic electron liquid interaction energies and static properties as computed from different dielectric formalism schemes with their \enquote{exact} counterparts as extracted from our PIMC simulations. It has been demonstrated that, beyond the warm dense matter regime and especially for $r_{\mathrm{s}}>30$, the STLS and the VS schemes yield increasingly inaccurate results\,\cite{HNCPIMC}. Only the IET, HNC and qSTLS schemes will be numerically solved herein, since inclusion of the STLS and VS schemes has been judged to be meaningless. The HNC and IET comparison will lead to direct conclusions regarding the impact of the classical OCP bridge function inclusion, while the qSTLS and IET comparison will lead to indirect conclusions regarding the significance of the beyond-RPA quantum effects. Our HNC algorithm can be essentially obtained from the IET algorithm described in Sec.\ref{numerical:algorithm} by setting $B(q)/[\beta{U}(q)]\equiv0$ and has been systematically validated against published results of Tanaka\,\cite{HNCSTLS}. On the other hand, our qSTLS algorithm has marked differences from the IET algorithm described in Sec.\ref{numerical:algorithm} owing to the dynamic nature of the LFC and has been benchmarked against published results of Schweng \& B\"ohm\,\cite{STLSqu2}.

\subsection{Interaction energy}\label{results:interactionenergy}

\noindent The interaction energy per particle is generally obtained by $U=(1/2)\int[d^3k/(2\pi)^3]{U}(\boldsymbol{k})\left[S(\boldsymbol{k})-1\right]$. Substituting for the Coulomb potential energy, introducing the $(r_{\mathrm{s}},\theta)$ variables and employing normalized wavenumbers, this leads to the standard UEF expression\,\cite{IchiRep,DornRev}
\begin{align}
\widetilde{u}(r_{\mathrm{s}},\theta)=\frac{1}{\pi\lambda{r}_{\mathrm{s}}}\int_0^{\infty}\left[S(x)-1\right]dx\,,\label{interactionenergies}
\end{align}
where $\widetilde{u}(r_{\mathrm{s}},\theta)$ denotes the interaction energy per particle normalized by the Hartree energy $E_{\mathrm{h}}=e^2/a_{\mathrm{B}}$.

The normalized interaction energies extracted from the PIMC simulations and computed with the qSTLS, HNC \& IET schemes have been listed in Table \ref{tab:internalenergies}. The absolute relative deviations between the theoretical and \enquote{exact} interaction energies have also been tabulated therein. \textbf{(i)} The qSTLS interaction energies are relatively inaccurate with $5.0\%-8.7\%$ relative deviations from PIMC results. The errors systematically decrease as $\theta$ increases and increase as $r_{\mathrm{s}}$ increases. \textbf{(ii)} The HNC interaction energies are very accurate with $0.47\%-1.37\%$ relative deviations from PIMC results. The errors systematically decrease as $\theta$ increases (diminishing quantum effects) and increase as $r_{\mathrm{s}}$ increases (stronger beyond-HNC classical pair correlations\,\cite{HNCbad1,HNCbad2,HNCbad3}). \textbf{(iii)} The IET interaction energies are revealed to be the most accurate with merely $0.05\%-0.68\%$ relative deviations from the PIMC results. No systematic tendencies have been observed in the errors with respect to either $\theta$ or $r_{\mathrm{s}}$. \textbf{(iv)} The IET scheme provides the most accurate interaction energy predictions for all $20$ thermodynamic state points investigated, with appreciable improvements over the HNC interaction energy predictions. \textbf{(v)} In spite of their very high accuracy, the IET (as well as the HNC) interaction energies cannot reproduce the exact $\theta$-dependence of $\widetilde{u}(r_{\mathrm{s}},\theta)$ within the highly degenerate $\theta\lesssim1$ range. To be more specific, the PIMC results reveal a monotonic $|\widetilde{u}|$ decrease as $\theta$ increases, while the IET (and HNC) results exhibit a monotonic $|\widetilde{u}|$ increase as $\theta$ increases within $\theta\lesssim1$ and the correct monotonic $|\widetilde{u}|$ decrease only as $\theta$ increases within $\theta\gtrsim1$. For this reason, we did not construct an exchange-correlation free energy parametrization via the adiabatic connection formula.

It is important to point out that the employed classical OCP bridge function parametrization is strictly valid for $10\leq\Gamma\leq170$\,\cite{BriOCP5,BriOCP6}. This upper threshold is surpassed by the state point $(r_{\mathrm{s}},\theta)=(200,0.50)$ which corresponds to $\Gamma=217.2$. Since all the $s_i(\Gamma)$ and $l_i(\Gamma)$ coefficients involved in Eq.(\ref{OCPbridgefunction}) are monotonic functions of $\Gamma$, it can be expected from the well-known continuity between the stable and metastable liquid states that the analytic classical OCP bridge function expression can be extrapolated without large errors towards the supercooled liquid regime $\Gamma>171.8$. The same applies for extrapolations in the weakly interacting regime $\Gamma<10$, but these are less significant due to the diminished bridge function impact.

\begin{table*}[t]
	\caption{The interaction energy $\widetilde{u}(r_{\mathrm{s}},\theta)$ per particle (expressed in Hartree units) of the paramagnetic electron liquid: comparison of the finite-size corrected \enquote{exact} PIMC results with the predictions of the qSTLS, HNC and IET dielectric formalism schemes. The PIMC data for the first $6$ state points are adopted from Ref.\cite{HNCPIMC}, while the PIMC data for the remaining $14$ state points are new. The absolute relative deviations between the dielectric scheme results and the PIMC results are also reported.}\label{tab:internalenergies}
	\centering
	\begin{tabular}{cccccccccc}
	\hline
	$r_{\mathrm{s}}$ & $\theta$     & $\Gamma$      & $\widetilde{u}_{\mathrm{PIMC}}$ & $\widetilde{u}_{\mathrm{qSTLS}}$ & $e_{\mathrm{qSTLS}}$ ($\%$) &  $\widetilde{u}_{\mathrm{HNC}}$ & $e_{\mathrm{HNC}}$ ($\%$) &  $\widetilde{u}_{\mathrm{IET}}$   &   $e_{\mathrm{IET}}$ ($\%$)  \\ \hline
    \,\,100\,\,      & \,\,0.50\,\, & \,\,108.6\,\, & \,\,-0.00825500\,\,                 & \,\,-0.00762382                      & 7.646                       & -0.00815866                     &     1.167                 & -0.00822181                       &        0.402                 \\
    \,\,100\,\,      & \,\,0.75\,\, & \,\,72.40\,\, & \,\,-0.00824570\,\,                 & \,\,-0.00764755                      & 7.254                       & -0.00816490                     &     0.980                 & -0.00822544                       &        0.246                 \\
    \,\,100\,\,      & \,\,1.00\,\, & \,\,54.30\,\, & \,\,-0.00823490\,\,                 & \,\,-0.00766712                      & 6.895                       & -0.00816618                     &     0.834                 & -0.00822559                       &        0.113                 \\
    \,\,100\,\,      & \,\,2.00\,\, & \,\,27.15\,\, & \,\,-0.00817650\,\,                 & \,\,-0.00768892                      & 5.963                       & -0.00812905                     &     0.580                 & -0.00819066                       &        0.173                 \\
    \,\,100\,\,      & \,\,4.00\,\, & \,\,13.58\,\, & \,\,-0.00800623\,\,                 & \,\,-0.00760308                      & 5.035                       & -0.00796833                     &     0.473                 & -0.00803143                       &        0.315                 \\
    \,\, 50\,\,      & \,\,0.50\,\, & \,\,54.30\,\, & \,\,-0.01600700\,\,                 & \,\,-0.01499807                      & 6.303                       & -0.01589841                     &     0.678                 & -0.01603510                       &        0.176                 \\
    \,\, 60\,\,      & \,\,0.50\,\, & \,\,65.16\,\, & \,\,-0.01345310\,\,                 & \,\,-0.01256041                      & 6.636                       & -0.01334804                     &     0.781                 & -0.01346014                       &        0.052                 \\
    \,\, 70\,\,      & \,\,0.50\,\, & \,\,76.02\,\, & \,\,-0.01161175\,\,                 & \,\,-0.01080732                      & 6.928                       & -0.01150938                     &     0.882                 & -0.01160390                       &        0.068                 \\
    \,\, 80\,\,      & \,\,0.50\,\, & \,\,86.88\,\, & \,\,-0.01021937\,\,                 & \,\,-0.00948545                      & 7.182                       & -0.01012012                     &     0.971                 & -0.01020149                       &        0.175                 \\
    \,\, 90\,\,      & \,\,0.50\,\, & \,\,97.74\,\, & \,\,-0.00912862\,\,                 & \,\,-0.00845287                      & 7.403                       & -0.00903293                     &     1.048                 & -0.00910415                       &        0.268                 \\
    \,\,110\,\,      & \,\,0.50\,\, & \,\,119.5\,\, & \,\,-0.00752642\,\,                 & \,\,-0.00694343                      & 7.746                       & -0.00744012                     &     1.147                 & -0.00749675                       &        0.394                 \\
    \,\,125\,\,      & \,\,0.50\,\, & \,\,135.8\,\, & \,\,-0.00665421\,\,                 & \,\,-0.00612435                      & 7.963                       & -0.00657377                     &     1.209                 & -0.00662268                       &        0.474                 \\
    \,\,125\,\,      & \,\,0.75\,\, & \,\,90.50\,\, & \,\,-0.00665053\,\,                 & \,\,-0.00614335                      & 7.626                       & -0.00657838                     &     1.085                 & -0.00662556                       &        0.442                 \\
    \,\,125\,\,      & \,\,1.00\,\, & \,\,67.88\,\, & \,\,-0.00664336\,\,                 & \,\,-0.00615966                      & 7.281                       & -0.00657999                     &     0.954                 & -0.00662647                       &        0.254                 \\
    \,\,125\,\,      & \,\,1.50\,\, & \,\,45.25\,\, & \,\,-0.00662535\,\,                 & \,\,-0.00617940                      & 6.731                       & -0.00657432                     &     0.770                 & -0.00662112                       &        0.064                 \\
    \,\,125\,\,      & \,\,2.00\,\, & \,\,33.94\,\, & \,\,-0.00660298\,\,                 & \,\,-0.00618476                      & 6.334                       & -0.00655900                     &     0.666                 & -0.00660712                       &        0.063                 \\
    \,\,150\,\,      & \,\,0.50\,\, & \,\,162.9\,\, & \,\,-0.00558177\,\,                 & \,\,-0.00511915                      & 8.288                       & -0.00550821                     &     1.318                 & -0.00554797                       &        0.606                 \\
    \,\,150\,\,      & \,\,1.00\,\, & \,\,81.45\,\, & \,\,-0.00557134\,\,                 & \,\,-0.00514888                      & 7.583                       & -0.00551337                     &     1.040                 & -0.00555132                       &        0.359                 \\
    \,\,200\,\,      & \,\,0.50\,\, & \,\,217.2\,\, & \,\,-0.00422244\,\,                 & \,\,-0.00385565                      & 8.687                       & -0.00416445                     &     1.373                 & -0.00419373                       &        0.680                 \\
    \,\,200\,\,      & \,\,1.00\,\, & \,\,108.6\,\, & \,\,-0.00421710\,\,                 & \,\,-0.00387821                      & 8.036                       & -0.00416813                     &     1.161                 & -0.00419559                       &        0.510                 \\  \hline
	\end{tabular}
\end{table*}

Furthermore, we compare with the interaction energies as computed from three accurate parametrizations of the exchange-correlation free energy through the expression
\begin{align}
\widetilde{u}(r_{\mathrm{s}},\theta)=r_{\mathrm{s}}\frac{\partial\widetilde{f}_{\mathrm{xc}}(r_{\mathrm{s}},\theta)}{\partial{r}_{\mathrm{s}}}+2\widetilde{f}_{\mathrm{xc}}(r_{\mathrm{s}},\theta)\,,\label{freeenergies}
\end{align}
that is acquired after the differentiation of the thermodynamic formula $\widetilde{f}_{\mathrm{xc}}(r_{\mathrm{s}},\Theta)=r^{-2}_{\mathrm{s}}\int_0^{r_{\mathrm{s}}}r_{\mathrm{s}}'\widetilde{u}_{\mathrm{int}}(r_{\mathrm{s}}',\Theta)dr_{\mathrm{s}}'$\,\cite{DornRev}. In particular, we consider the  parametrization by Groth \emph{et al.} (GDSMFB) that is based on simulation results obtained by various novel PIMC methods within $0.1\leq{r}_{\mathrm{s}}\leq20$ \& $0.5\leq\theta\leq8$\,\cite{parame1}, the parametrization by Karasiev \emph{et al.} (KSDT) that is based on simulation results obtained by the restricted PIMC method within $1.0\leq{r}_{\mathrm{s}}\leq40$ \& $0.0625\leq\theta\leq8$\,\cite{parame2} and the corrected version of the parametrization by Karasiev \emph{et al.} (corrKSDT) that is also based on the above restricted PIMC data\,\cite{parame3}. Despite some documented deficiencies of the restricted PIMC data input\,\cite{parame4} (concerning the uncontrolled fixed node approximation\,\cite{parame5} and the unsatisfactory treatment of finite-size effects\,\cite{parame6}) and in spite of a procedural mistake in the original fitting procedure (concerning the utilization of an analytic ground state fit instead of the actual ground state MC data\,\cite{parame7}), the KSDT interaction energies exhibit $0.39\%-1.63\%$ relative deviations from PIMC results, whereas the GDSMFB interaction energies exhibit $1.08\%-4.94\%$ relative deviations and the corrKSDT interaction energies exhibit $1.58\%-6.90\%$ relative deviations from PIMC results. Therefore, extrapolated GDSMFB, KSDT and corrKSDT interaction energies are more accurate than the qSTLS results, but less accurate than the IET results. More specifically, the KSDT interaction energies are even more accurate than the HNC results at some investigated states, but never more accurate than the IET results. Hence, even though the GDSMFB and corrKSDT exchange-correlation free energy parametrization is much more accurate than the KSDT parametrization within the warm dense matter regime\,\cite{parame1}, it can be concluded that the latter extrapolates better within the strongly coupled electron regime, at least as far as interaction energies are concerned. For completeness, we point out that the GDSMFB, KSDT, corrKSDT discrepancies in warm dense matter ranges do not impact density functional theory calculations\,\cite{parame8}.

Finally, we also compare with the interaction energies as computed from the direct parametrization of the interaction energy by Ichimaru \emph{et al.} (IIT) which is based on the STLS interaction energies after their correction, to comply with variable coupling QMC simulations for the ground state limit ($\theta\to0$) and with variable coupling MC simulations for the classical limit ($\theta\to\infty$), via the implementation of an ad hoc $\theta-$interpolation function\,\cite{IchiRep,parame9,param10}. Although the interpolation function accuracy is unclear for intermediate degeneracies\,\cite{DornRev}, the incorporation of exact strong coupling ground state results\,\cite{param11} and classical results\,\cite{param12,param13} hints that the IIT parametrization might be very accurate. This is verified by the comparison which reveals that the IIT interaction energies exhibit $0.06\%-0.63\%$ relative deviations from PIMC results. Remarkably, IIT interaction energies are much more accurate than the qSTLS, more accurate than the HNC and even as accurate as the IET results. In particular, the IET interaction energies are more accurate for $11$ and the IIT interaction energies more accurate for $9$ of the studied states, while the mean absolute relative errors are $0.29\%$ for the IET, $0.35\%$ for the IIT.

\subsection{Static structure factor}\label{results:structurefactor}

\begin{figure}
	\centering
	\includegraphics[width=3.40in]{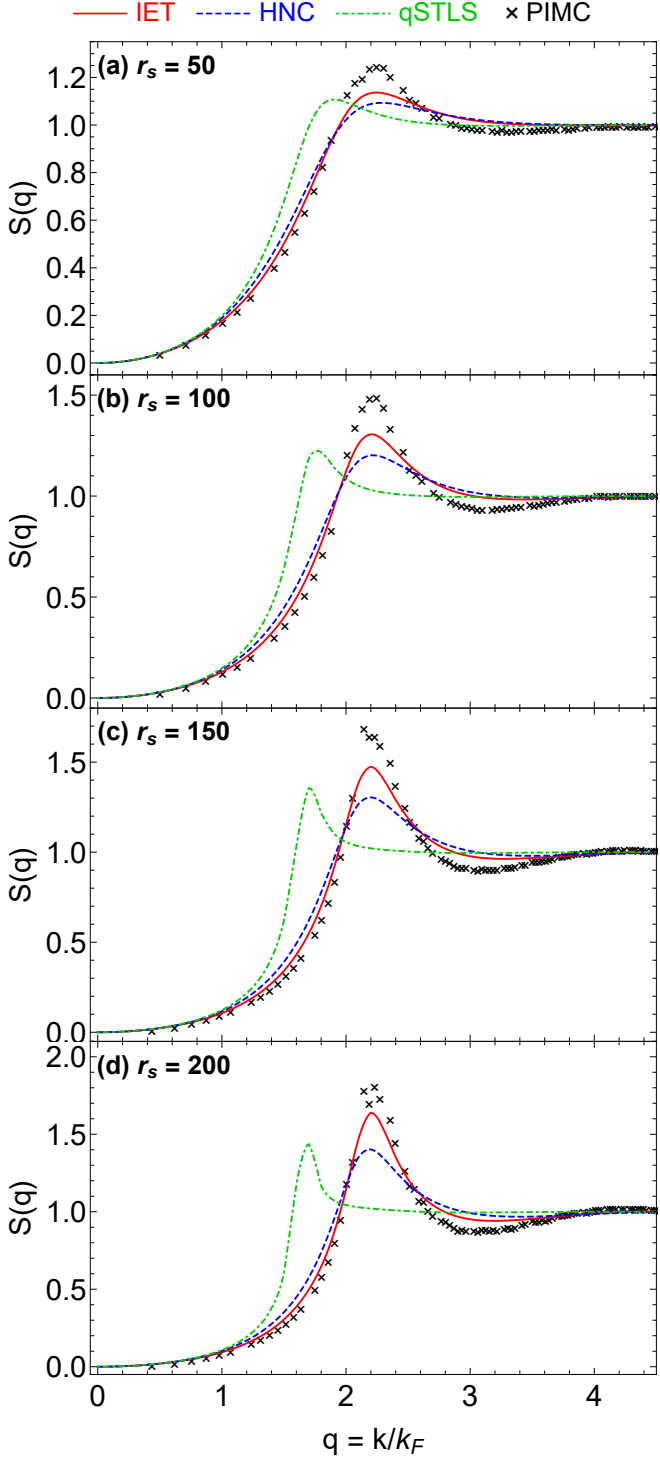}
	\caption{Dependence of the paramagnetic electron liquid static structure factor on the quantum coupling parameter $r_{\mathrm{s}}$. Results from the IET scheme (red solid line), the HNC scheme (blue dashed line), the qSTLS scheme (green dot-dashed line) and PIMC simulations (black crosses) for $\theta=0.50$ and varying $r_{\mathrm{s}}=50,100,150,200$ in the normalized wavenumber range $k\leq4.5{k}_{\mathrm{F}}$. The superiority of the new IET scheme within the long wavelength range, in the maximum vicinity and within the short wavelength range is obvious.}\label{fig:ssfcoupling}
\end{figure}

\begin{figure}
	\centering
	\includegraphics[width=3.40in]{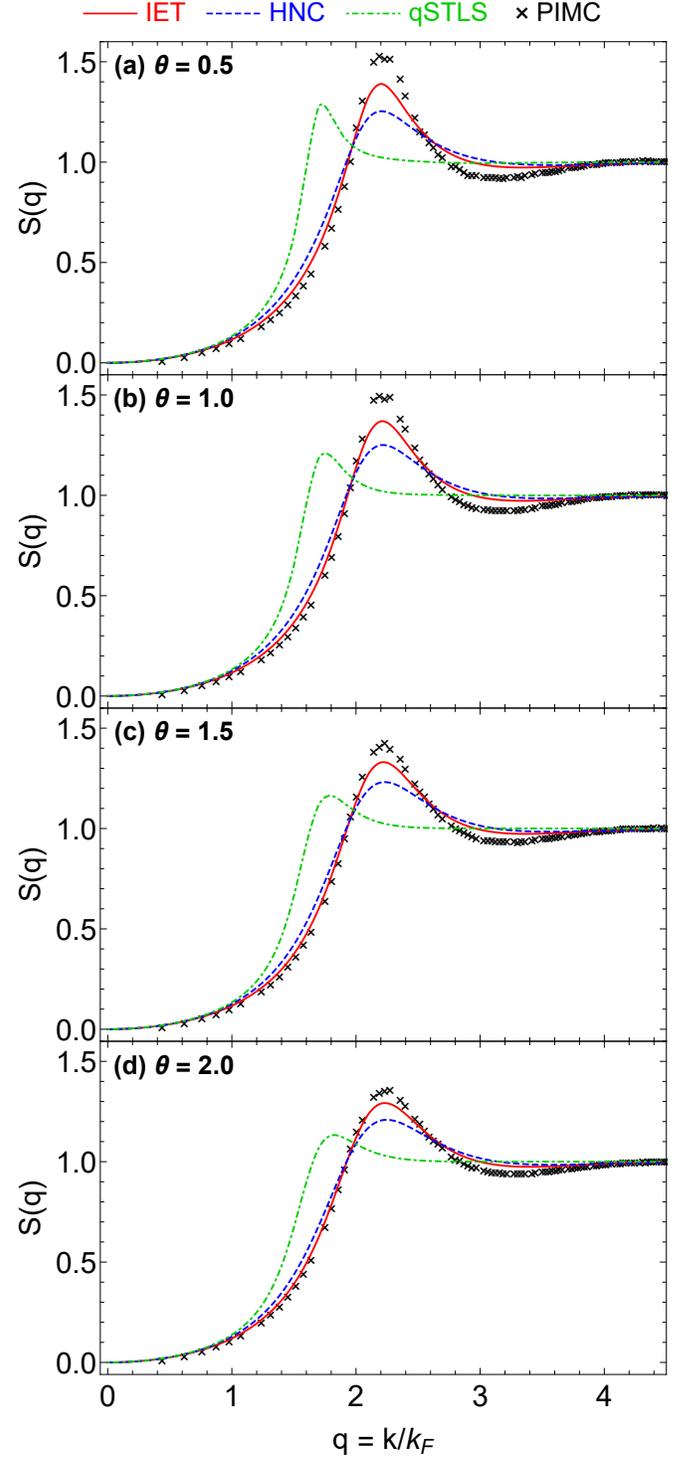}
	\caption{Dependence of the paramagnetic electron liquid static structure factor on the quantum degeneracy parameter $\theta$. Results from the IET scheme (red solid line), the HNC scheme (blue dashed line), the qSTLS scheme (green dot-dashed line) and PIMC simulations (black crosses) for $r_{\mathrm{s}}=125$ and varying $\theta=0.5,1.0,1.5,2.0$ in the normalized wavenumber range $k\leq4.5{k}_{\mathrm{F}}$. The superiority of the new IET scheme within the long wavelength range, in the maximum vicinity and within the short wavelength range is obvious.}\label{fig:ssfdegeneracy}
\end{figure}

\noindent Characteristic static structure factors extracted from our PIMC simulations and computed with the three dielectric formalism schemes have been illustrated in Fig.\ref{fig:ssfcoupling} and Fig.\ref{fig:ssfdegeneracy}. For all the $20$ investigated state points, the magnitudes and positions of the first SSF peak resulting from the PIMC simulations as well as from the qSTLS, HNC \& IET schemes have been listed in Table \ref{tab:structurefactors}. The absolute relative deviations between the theoretical and the \enquote{exact} SSF peak values have also been tabulated therein.

\begin{table*}[t]
	\caption{The peak magnitude ${S}^{\mathrm{max}}$ and peak position ${\arg}_q{S}^{\mathrm{max}}$ of the static structure factor $S(k/k_{\mathrm{F}};r_{\mathrm{s}},\theta)$ of the paramagnetic electron liquid: comparison of the \enquote{exact} PIMC results with the predictions of the qSTLS, HNC and IET dielectric formalism schemes. The PIMC data for the first $6$ state points are adopted from Ref.\cite{HNCPIMC}, while the PIMC data for the remaining $14$ state points are new. The absolute relative deviations between the dielectric scheme results and the PIMC results are also reported.}\label{tab:structurefactors}
	\centering
	\begin{tabular}{cc|ccccccc|ccccccc}
	\hline
	$r_{\mathrm{s}}$ & $\theta$ & ${S}^{\mathrm{max}}_{\mathrm{PIMC}}$ & ${S}^{\mathrm{max}}_{\mathrm{qSTLS}}$ & $e_{\mathrm{qSTLS}}$ & ${S}^{\mathrm{max}}_{\mathrm{HNC}}$ & $e_{\mathrm{HNC}}$ & ${S}^{\mathrm{max}}_{\mathrm{IET}}$ & $e_{\mathrm{IET}}$ &
$\underset{q}{\arg}{S}^{\mathrm{max}}_{\mathrm{PIMC}}$ & $\underset{q}{\arg}{S}^{\mathrm{max}}_{\mathrm{qSTLS}}$ & $e_{\mathrm{qSTLS}}$ & $\underset{q}{\arg}{S}^{\mathrm{max}}_{\mathrm{HNC}}$ & $e_{\mathrm{HNC}}$ & $\underset{q}{\arg}{S}^{\mathrm{max}}_{\mathrm{IET}}$ & $e_{\mathrm{IET}}$ \\
	&  &  &  & ($\%$) &  & ($\%$) &  &  ($\%$) &  &  & ($\%$) &  & ($\%$) &  & ($\%$) \\ \hline
100  & 0.50 & 1.494 	& 1.224 	& 18.074    & 1.202 	& 19.546 	& 1.306 	& 12.598 	& 2.248		& 1.770 	& 21.247 	& 2.220 	& 1.225 	& 2.200 	& 2.115          \\
100  & 0.75 & 1.449 	& 1.191 	& 17.779    & 1.203 	& 16.932 	& 1.301 	& 10.159 	& 2.248		& 1.770 	& 21.247 	& 2.220 	& 1.225 	& 2.210 	& 1.670          \\
100  & 1.00 & 1.406 	& 1.166 	& 17.127    & 1.199 	& 14.724 	& 1.290 	& 8.285  	& 2.191		& 1.790 	& 18.288 	& 2.230 	& 1.797 	& 2.220 	& 1.341          \\
100  & 2.00 & 1.285 	& 1.102 	& 14.204    & 1.164 	& 9.366  	& 1.227 	& 4.461  	& 2.248		& 1.870	    & 16.798 	& 2.270 	& 0.999 	& 2.250 	& 0.110          \\
100  & 4.00 & 1.166 	& 1.051 	& 9.912     & 1.105 	& 5.263  	& 1.138 	& 2.381  	& 2.303		& 2.040 	& 11.422 	& 2.340 	& 1.605 	& 2.320 	& 0.736          \\
 50  & 0.50 & 1.252 	& 1.107 	& 11.569    & 1.093 	& 12.677 	& 1.136 	& 9.211  	& 2.248		& 1.910 	& 15.018 	& 2.290 	& 1.889 	& 2.250 	& 0.110          \\
 60  & 0.50 & 1.327 	& 1.131 	& 14.767    & 1.115 	& 15.956 	& 1.171 	& 11.802 	& 2.231		& 1.860 	& 16.635 	& 2.260 	& 1.294 	& 2.230 	& 0.051          \\
 70  & 0.50 & 1.351 	& 1.156 	& 14.450    & 1.137 	& 15.822 	& 1.205 	& 10.861 	& 2.188		& 1.830 	& 16.355 	& 2.250 	& 2.843 	& 2.220 	& 1.471          \\
 80  & 0.50 & 1.391 	& 1.180 	& 15.174    & 1.159 	& 16.674 	& 1.238 	& 10.990 	& 2.144		& 1.800 	& 16.029 	& 2.230 	& 4.030 	& 2.210 	& 3.097          \\
 90  & 0.50 & 1.438 	& 1.202 	& 16.418    & 1.181 	& 17.905 	& 1.272 	& 11.551 	& 2.231		& 1.790 	& 19.772 	& 2.220 	& 0.499 	& 2.210 	& 0.947          \\
110  & 0.50 & 1.481 	& 1.248 	& 15.766    & 1.223 	& 17.456 	& 1.339 	& 9.591  	& 2.144		& 1.750 	& 18.362 	& 2.210 	& 3.097 	& 2.200 	& 2.631          \\
125  & 0.50 & 1.538 	& 1.288 	& 16.310    & 1.254 	& 18.502 	& 1.390 	& 9.661  	& 2.188		& 1.720 	& 21.383 	& 2.200 	& 0.557 	& 2.200 	& 0.557          \\
125  & 0.75 & 1.557 	& 1.243 	& 20.160    & 1.256 	& 19.369 	& 1.384 	& 11.132 	& 2.231		& 1.730 	& 22.461 	& 2.210 	& 0.947 	& 2.200 	& 1.395          \\
125  & 1.00 & 1.501 	& 1.209 	& 19.455    & 1.251 	& 16.671 	& 1.369	 	& 8.765     & 2.188		& 1.750 	& 20.011 	& 2.220 	& 1.471 	& 2.210 	& 1.014          \\
125  & 1.50 & 1.436 	& 1.163 	& 19.000    & 1.231 	& 14.258 	& 1.331	 	& 7.338     & 2.231		& 1.790 	& 19.772 	& 2.230 	& 0.051 	& 2.220 	& 0.499          \\
125  & 2.00 & 1.365 	& 1.133 	& 17.039    & 1.209 	& 11.464 	& 1.292	 	& 5.360     & 2.274		& 1.820 	& 19.952 	& 2.250 	& 1.040 	& 2.230 	& 1.919          \\
150  & 0.50 & 1.695 	& 1.355 	& 20.022    & 1.305 	& 23.018 	& 1.474	    & 13.022    & 2.144		& 1.700 	& 20.694	& 2.200 	& 2.631 	& 2.200 	& 2.631          \\
150  & 1.00 & 1.586 	& 1.255 	& 20.916    & 1.300 	& 18.038 	& 1.447	 	& 8.768     & 2.231		& 1.720 	& 22.909 	& 2.210 	& 0.947 	& 2.210 	& 0.947          \\
200  & 0.50 & 1.817 	& 1.436 	& 20.962    & 1.403 	& 22.807 	& 1.638	 	& 9.833     & 2.231		& 1.700 	& 23.806 	& 2.190 	& 1.844 	& 2.200 	& 1.395          \\
200  & 1.00 & 1.738 	& 1.337 	& 23.112    & 1.395 	& 19.757 	& 1.600	 	& 7.943     & 2.231		& 1.700 	& 23.806 	& 2.200 	& 1.395 	& 2.200 	& 1.395          \\  \hline
	\end{tabular}
\end{table*}

For the state points of interest, the liquid character of the UEF becomes apparent from the relatively large magnitude of the first SSF peak (which is generally around $\sim1.5$ and even reaches $1.82$), the well-resolved first SSF trough around $3k_{\mathrm{F}}$ (especially for $r_{\mathrm{s}}\gtrsim100$) and the well-resolved second SSF peak above $4k_{\mathrm{F}}$ (only for $r_{\mathrm{s}}\gtrsim100$). \textbf{(i)} The IET scheme always generates the most accurate SSF across the entire interval: within the long wavelength range of $k\lesssim2k_{\mathrm{F}}$, in the Lorentzian shaped region that surrounds the first maximum $k\sim2-2.5k_{\mathrm{F}}$ and within the short wavelength range of $k\gtrsim2.5k_{\mathrm{F}}$. \textbf{(ii)} The qSTLS scheme has the worst performance. It strongly underestimates the position of the first SSF peak by around $20\%$. The same behavior was earlier observed for the STLS and the VS scheme\,\cite{HNCPIMC}. These results can be anticipated from classical OCP liquids, where it has been established that stronger Coulomb correlations (neglected in STLS, VS and qSTLS) not only increase the SSF peak magnitude but also slightly displace it towards longer wavenumbers. \textbf{(iii)} On the other hand, the IET and HNC schemes provide very accurate predictions for the SSF peak positions, $0.05\%-3.10\%$ and $0.05\%-4.03\%$ relative deviations from the PIMC results respectively, with the IET scheme having the slight edge. Thus, it can concluded that the first SSF peak position is mainly controlled by strong correlations. \textbf{(iv)} Regardless of state, the IET scheme greatly improves the HNC prediction for the SSF peak magnitude, with the relative deviations from PIMC results being $2.38\%-13.02\%$ and $5.26\%-23.02\%$, respectively. \textbf{(v)} For all states, the IET SSF is characterized by a very accurate long wavelength behavior, especially for $k\lesssim{k}_{\mathrm{F}}$.

Despite the large improvements over the HNC scheme, the IET scheme seems to generate SSFs that are not accurate enough to justify the very accurate predictions of interaction energies. Detailed inspection of Figs.\ref{fig:ssfcoupling},\ref{fig:ssfdegeneracy} reveals that the very accurate interaction energies are the result of favorable error cancellation in Eq.(\ref{interactionenergies}), since the IET SSF tends to be slightly too large within $k_{\mathrm{F}}\leq{k}\leq2k_{\mathrm{F}}$, becomes clearly too small within $2.0k_{\mathrm{F}}\leq{k}\leq2.5k_{\mathrm{F}}$ and tends to be somewhat too large within $2.5k_{\mathrm{F}}\leq{k}\leq4.0k_{\mathrm{F}}$. The same reasoning applies for the HNC scheme. Note that a similar favorable error cancellation is responsible for the success of STLS generated interaction energies within the warm dense matter regime\,\cite{DornRev,param10}.

\begin{figure}
	\centering
	\includegraphics[width=3.40in]{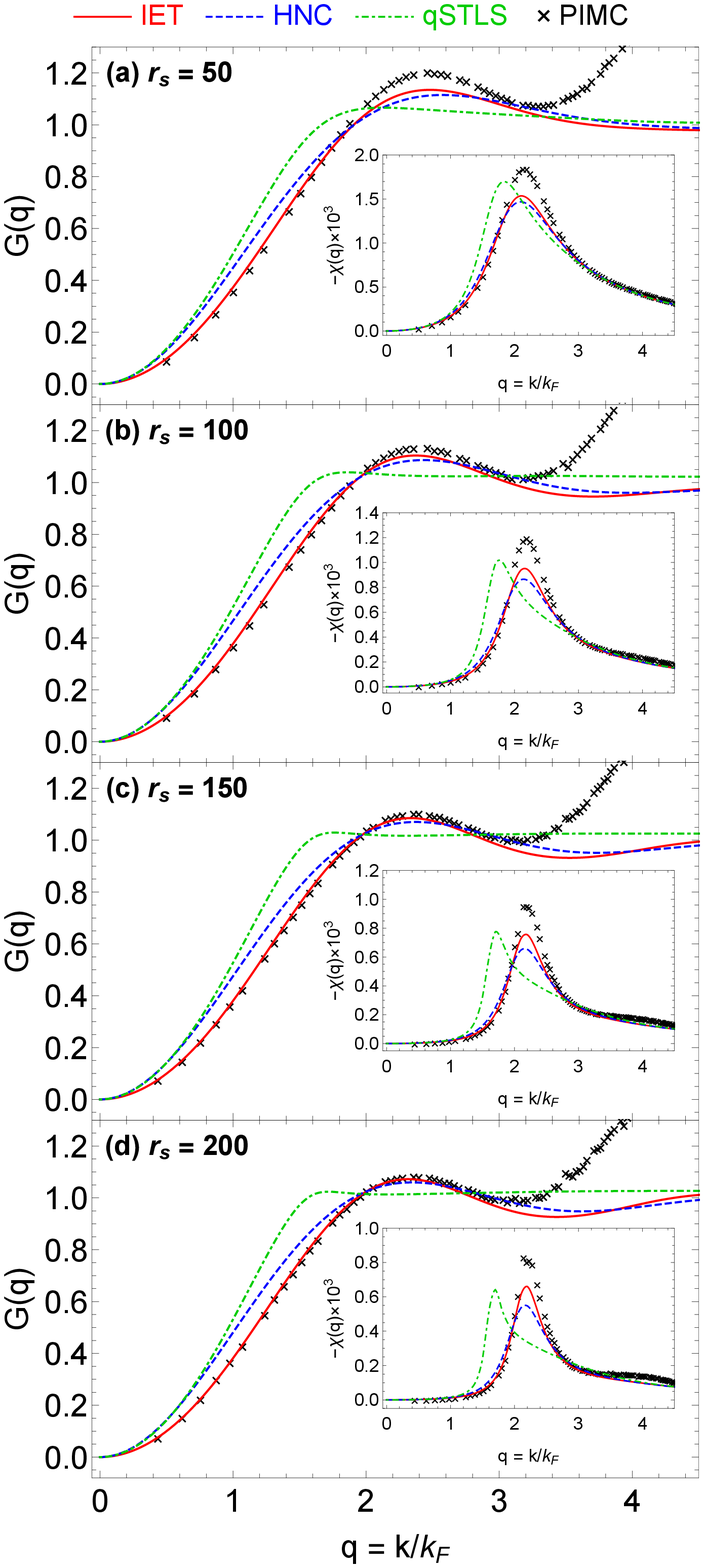}
	\caption{Dependence of the paramagnetic electron liquid static local field correction (main figure) and static density response (inset figure) on the quantum coupling parameter $r_{\mathrm{s}}$. Results from the IET scheme (red solid line), the HNC scheme (blue dashed line), the qSTLS scheme (green dot-dashed line) and our PIMC simulations (black crosses) for $\theta=1.00$ and varying $r_{\mathrm{s}}=50,100,150,200$ in the normalized wavenumber range $k\leq4.5{k}_{\mathrm{F}}$. For both quantities, the superiority of the new IET scheme within the long wavelength range and in the vicinity of the maximum is obvious.}\label{fig:lfccoupling}
\end{figure}

\begin{figure}
	\centering
	\includegraphics[width=3.40in]{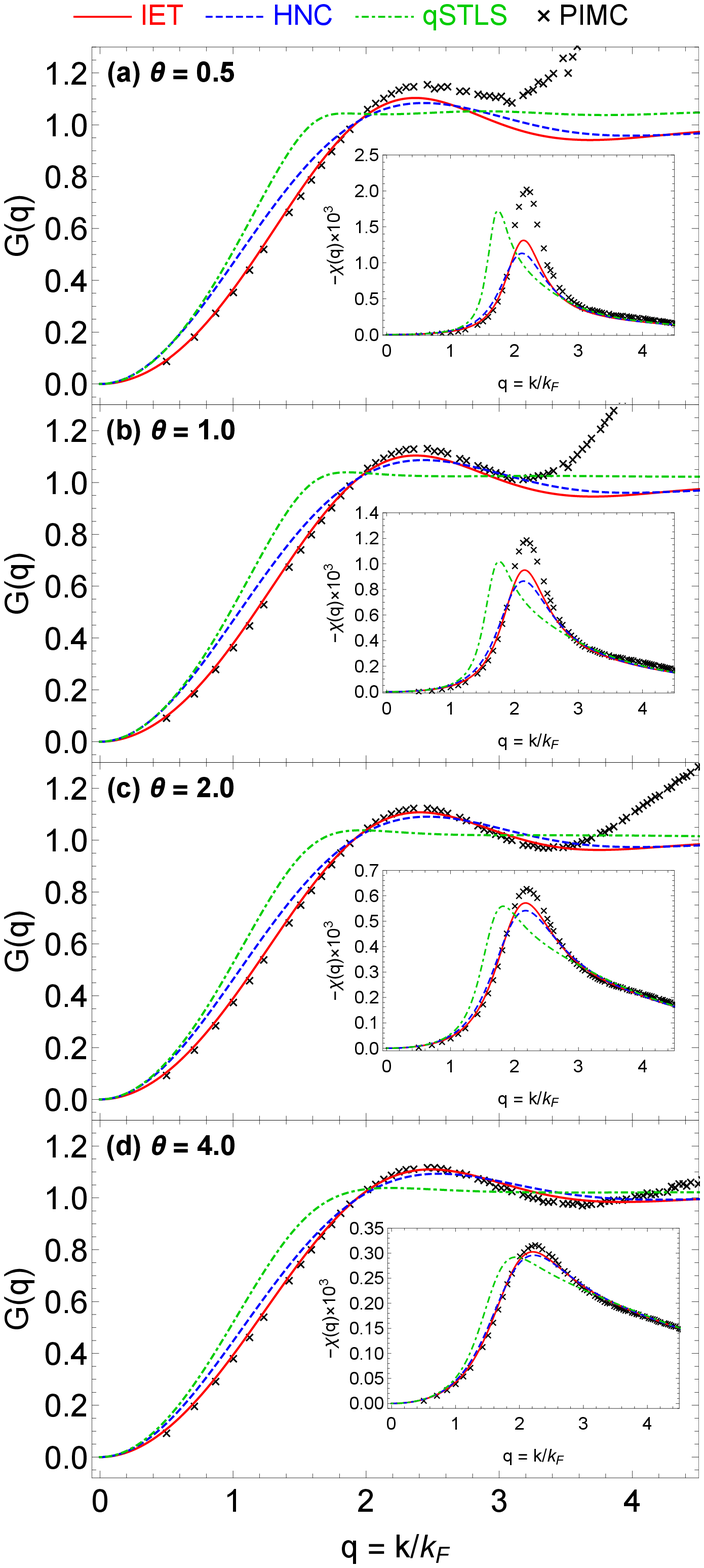}
	\caption{Dependence of the paramagnetic electron liquid static local field correction (main figure) and static density response (inset figure) on the quantum degeneracy parameter $\theta$. Results from the IET scheme (red solid line), the HNC scheme (blue dashed line), the qSTLS scheme (green dot-dashed line) and our PIMC simulations (black crosses) for $r_{\mathrm{s}}=100$ and varying $\theta=0.5,1,2,4$ in the normalized wavenumber range $k\leq4.5{k}_{\mathrm{F}}$. For both quantities, the superiority of the new IET scheme within the long wavelength range and in the vicinity of the maximum is obvious.}\label{fig:lfcdegeneracy}
\end{figure}

Let us briefly discuss the radial distribution functions $g(r)$ (RDFs) extracted from PIMC simulations and computed with the three dielectric formalism schemes. Since RDFs encode the same correlation information as SSFs, it is unsurprising that the IET scheme greatly improves the predictions of the HNC scheme for the first maximum and first non-zero minimum magnitude, that the IET \& HNC schemes provide very accurate similar predictions for the first maximum and first non-zero minimum positions and that the qSTLS scheme has the worst performance. Finally, it is important to discuss the un-physical negative RDF region near the origin; a common pathology of all dielectric formalism schemes stemming from the approximate treatment of quantum effects\,\cite{DornRev}. The IET \& HNC schemes have negative regions of similar extent and signal power, as expected from the RPA-level treatment of quantum effects. On the other hand, the qSTLS scheme has a negative region of less extent and power, as expected from its beyond-RPA quantum features.

\subsection{Static local field correction and static density response}\label{results:localfieldcorrection}

\noindent Characteristic static local field corrections and static density responses extracted from our PIMC simulations and computed with the three dielectric formalism schemes have been illustrated in Fig.\ref{fig:lfccoupling} and Fig.\ref{fig:lfcdegeneracy}.

Let us first discuss the \emph{static local field correction}:\,\textbf{(i)} The PIMC LFC exhibits a pronounced first local maximum of magnitude $\sim1.10-1.25$ located at $k\sim2.35-2.60k_{\mathrm{F}}$, which is followed by a shallow local minimum located at $k\sim3.05-3.45k_{\mathrm{F}}$ and is ultimately succeeded by a rather sharp short wavelength increase. The asymptotic behavior does not contradict the short wavelength limit $G(k\to\infty)=1-g(0)$\,\cite{IchiMat,asympt3} that was discussed earlier, since this is valid for frequency independent LFCs (as confirmed by the HNC and the IET asymptotics), but not for the exact static LFC. In the ground state $\theta\to0$, the asymptotic behavior of the exact static LFC is described by the Holas expansion $G(k\to\infty,0)=B+Ck^2$ which predicts a parabolic divergence\,\cite{asympt5}. Such a parabolic asymptote has been empirically observed to persist also when $\theta\sim1$\,\cite{HNCPIMC,asympt6}, but not in the classical limit $\theta\gg1$, where one gets $G(k\to\infty,0)=1$\,\cite{IchiRep}. The above observation is confirmed by the present PIMC LFC, since, as $\theta$ increases, the asymptotic limit gradually switches from parabolically diverging to nearly unity in the depicted wavenumber range, see Fig.\ref{fig:lfcdegeneracy}(a-d). \textbf{(ii)} Regardless of the state point of interest, the IET scheme generates the most accurate and the qSTLS scheme the least accurate static LFC. The IET scheme improves the HNC prediction for the LFC first peak magnitude, with the relative deviations from PIMC results being $1.4\%-10.1\%$ and $2.6\%-12.0\%$, respectively. However, the HNC prediction for the LFC first peak position is always more accurate. \textbf{(iii)} Perhaps, the most remarkable feature of the IET scheme's LFC concerns its very accurate long wavelength behavior that extends up to nearly $k\lesssim2{k}_{\mathrm{F}}$. This does not necessarily guarantee that the IET scheme satisfies to a high degree the compressibility sum rule (CSR), an exact relation connecting the long wavelength static LFC to the second density derivative of the exchange-correlation free energy, \emph{i.e.} in $(n,T)$ thermodynamic variables and cgs units\,\cite{IchiRep,DornRev,IchiRev,Ichibok}
\begin{equation}
\lim_{k\to0}\frac{G(k)}{k^2}=-\frac{1}{4\pi{e}^2}\frac{\partial^2}{\partial{n}^2}\left[nf_{\mathrm{ex}}(n,T)\right]\,,\label{CSRnatural}
\end{equation}
or in $(r_{\mathrm{s}},\theta)$ thermodynamic variables and Hartree units
\begin{align}
\lim_{x\to0}\frac{G(x)}{x^2}&=-\frac{\pi}{12}\lambda{r}_{\mathrm{s}}\left(4\Theta^2\frac{\partial^2}{\partial{\Theta}^2}+r_{\mathrm{s}}^2\frac{\partial^2}{\partial{r_{\mathrm{s}}^2}}+4\Theta{r}_{\mathrm{s}}\frac{\partial^2}{\partial{\Theta}\partial{r_{\mathrm{s}}}}\right.\nonumber\\&\quad\qquad\left.-2\Theta\frac{\partial}{\partial{\Theta}}-2r_{\mathrm{s}}\frac{\partial}{\partial{r_{\mathrm{s}}}}\right)\left[\widetilde{f}_{\mathrm{xc}}(r_{\mathrm{s}},\theta)\right]\,.\label{CSRnorm}
\end{align}
This is simply because each dielectric scheme satisfies its individual CSR that involves its individual exchange-correlation free energy and not the exact exchange-correlation free energy. However, when combined with the very accurate IET interaction energies, the very accurate IET LFC long wavelength limit makes it very likely that the IET scheme satisfies the CSR to a large degree. We postpone such an investigation to a future work owing to its high computational cost, since as Eq.(\ref{CSRnorm}) suggests, it requires an accurate parametrization of the IET exchange-correlation free energy. Nevertheless, this IET feature can be exploited to provide an alternative route for the accurate calculation of the paramagnetic electron liquid's isothermal compressibility without involving neither PIMC simulations nor thermodynamic integration.

Let us now discuss the \emph{static density response function} defined by $\chi(\boldsymbol{k})\equiv\chi(\boldsymbol{k},\omega=0)$. \textbf{(i)} The qualitative behavior of the \enquote{exact} $\chi(\boldsymbol{k})$ of the paramagnetic electron liquid has been discussed earlier\,\cite{HNCPIMC} and is confirmed by our PIMC results. It is evident that the magnitude of the pronounced $\chi(\boldsymbol{k})$ extremum increases as $r_{\mathrm{s}}$ decreases (see Fig.\ref{fig:lfccoupling}) and as $\theta$ decreases (see Fig.\ref{fig:lfcdegeneracy}) and that the width of the $\chi(\boldsymbol{k})$ extremum becomes more sharp as $r_{\mathrm{s}}$ increases (see Fig.\ref{fig:lfccoupling}) and especially as $\theta$ decreases (see Fig.\ref{fig:lfcdegeneracy}). \textbf{(ii)} Since $\chi(\boldsymbol{k})$ is obtained by setting $\omega=0$ to Eq.(\ref{densityresponseDLFC}) and only involves the common among schemes $\chi_0(\boldsymbol{k})$ and the varying among schemes $G(\boldsymbol{k})$, comparison with respect to $\chi(\boldsymbol{k})$ should reflect the comparison with respect to $G(\boldsymbol{k})$. In fact, the IET scheme generates the most accurate $\chi(\boldsymbol{k})$ and the qSTLS scheme the least accurate $\chi(\boldsymbol{k})$. The IET scheme strongly improves the HNC prediction for the extremum magnitude, but the HNC prediction for the extremum position is somewhat more accurate. \textbf{(iii)} Finally, the IET scheme's static density response is again nearly exact for long wavelengths up to nearly $k\lesssim2{k}_{\mathrm{F}}$.

\begin{figure*}
	\centering
	\includegraphics[width=6.70in]{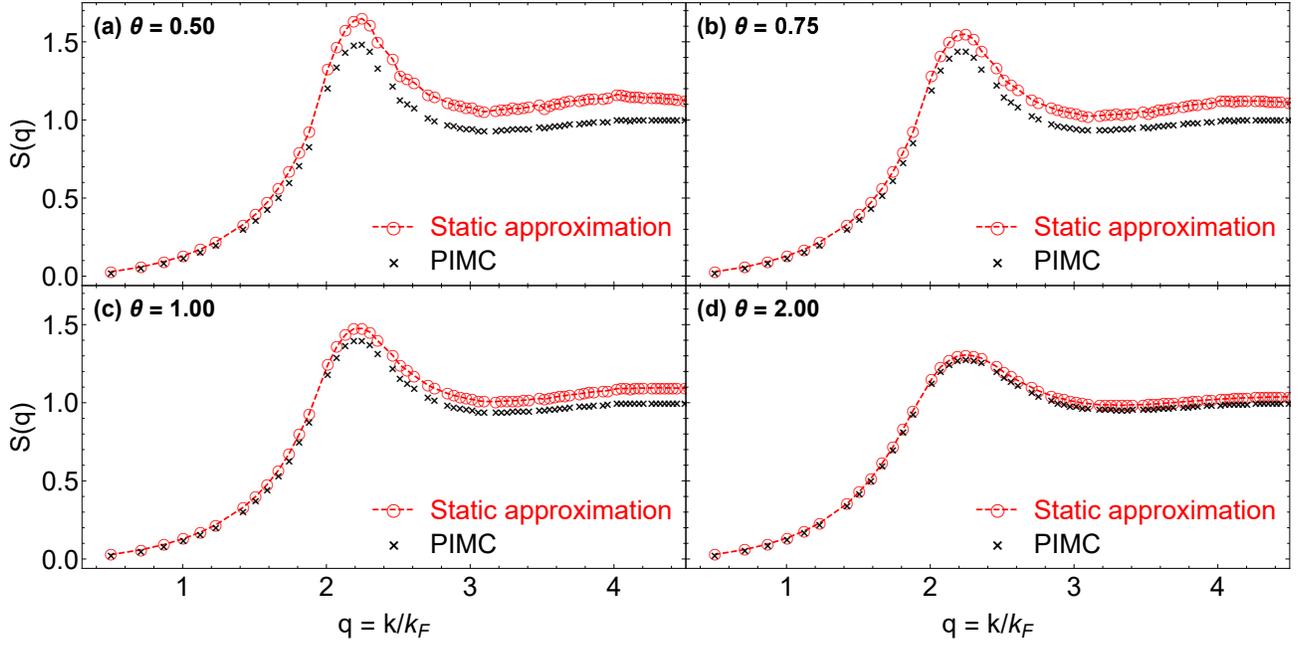}
	\caption{The paramagnetic electron liquid's static structure factor for $r_{\mathrm{s}}=100$ and $\theta=0.50$ (a), $\theta=0.75$ (b), $\theta=1.00$ (c), $\theta=2.00$ (d). Results from the static approximation (red solid line and red circles) and PIMC simulations (black crosses) in the normalized wavenumber range $k\leq4.5{k}_{\mathrm{F}}$. The static approximation consistently overestimates the SSF, especially at small values of the degeneracy parameter, which leads to highly inaccurate interaction energies.}\label{fig:SAexamples}
\end{figure*}

\subsection{Static approximation and effective static approximation}\label{results:SA&ESAextrapolation}

\noindent The \emph{static approximation} utilizes the exact static LFC PIMC results to close the set of equations of the dielectric formalism\,\cite{intro29,static2}. In other words, this approach replaces the exact dynamic LFC $G(\boldsymbol{k},\omega)$ with its exact static limit $G(\boldsymbol{k})=G(\boldsymbol{k},\omega=0)$ that is accessible from PIMC simulations. The approach has been demonstrated to yield basically exact results for the dynamic structure factor and other related dynamic quantities in their entire non-trivial range provided that $r_{\mathrm{s}}\lesssim4$ and to even reproduce the most prominent characteristics of the same quantities even at stronger coupling\,\cite{static2}. The approach has also been revealed to lead to accurate results for the SSF only for $k\lesssim2.5k_{\mathrm{F}}$ but to overestimate short-range correlations; a shortcoming which stems from the divergence of the exact static LFC short wavelength limit\,\cite{ESAsPRL}. The SSF-related accuracy of the static approximation is expected to deteriorate further at strong coupling, where the dynamic nature of the LFC has been documented to be more important even in the classical limit\,\cite{Ichibok,ClaCor3,ClaCor4}. In order to draw more quantitative conclusions, the exact static LFC PIMC results were employed as the closure of the exact two building blocks of the dielectric formalism, see Eqs.(\ref{Matsubaraseries},\ref{densityresponseDLFC}). It is evident that the static approximation has a low computational cost similar to that of the RPA.

Systematic comparisons revealed the following: \textbf{(i)} Regardless of the state point, the static approximation begins to overestimate the SSF prior to the main peak. For some state points, the overestimation takes place already at $k\lesssim1.5k_{\mathrm{F}}$, \emph{i.e.} right after the long wavelength range. \textbf{(ii)} As expected, the static approximation is much less accurate in the strongly coupled regime than in the warm dense matter regime. However, as $\theta$ increases, the accuracy level becomes significantly higher and the deviations become restricted at progressively shorter wavelengths, as revealed in Fig.\ref{fig:SAexamples}. \textbf{(iii)} Since the static approximation either reproduces or overestimates the SSF in the entire wavenumber range, there is no possibility of favorable error cancellation during the computation of the interaction energy. As a result, the absolute relative deviations in the interaction energies always exceed $10\%$ and can even reach $40\%$.

Moreover, given the success of the IIT interaction energy parametrization in the strongly coupled regime (see section \ref{results:interactionenergy}), it is tempting to check whether an extrapolation of the \emph{effective static approximation} (ESA)\,\cite{ESAsPRL,ESAsPRB} scheme would also prove to be successful. The ESA utilizes an analytical representation for the frequency independent LFC\,\cite{ESAsPRB} which is based on a neural net representation of PIMC static LFC results\,\cite{asympt6}, but also incorporates the static LFC exact long wavelength limit as expressed by the CSR (employing the GDSMFB exchange-correlation free energy parametrization\,\cite{parame1}) and the frequency independent LFC exact short wavelength limit as expressed by the asymptotic self-consistency condition (employing a $g(0)$ parametrization based on extrapolated restricted PIMC data\,\cite{ESAsPRL}). Owing to the latter feature, the ESA scheme does not suffer from the aforementioned drawback of the static approximation. In fact, the ESA was recently demonstrated to be a fast and reliable tool for the computation of numerous thermodynamic, static and dynamic quantities within the warm dense matter relevant range of $0.7\leq{r}_{\mathrm{s}}\leq20$ and $0\leq\theta\leq4$\,\cite{ESAsPRL,ESAsPRB}. Taking into account the satisfactory performance of the extrapolation of the GDSMFB parametrization towards low densities as well as the very small (albeit unphysical negative) values of the extrapolated $g(0)$ parametrization towards low densities, the ESA performance for the paramagnetic electron liquid mainly depends on whether the analytical representation of the intermediate wavelength PIMC static LFC results can be successfully extrapolated towards high coupling parameters. In order to draw conclusions, the analytical ESA LFC was used as the closure of the exact two building blocks of the dielectric formalism, see Eqs.(\ref{Matsubaraseries},\ref{densityresponseDLFC}). It is evident that the ESA scheme also has a low computational cost similar to that of the RPA.

\begin{figure*}
	\centering
	\includegraphics[width=6.70in]{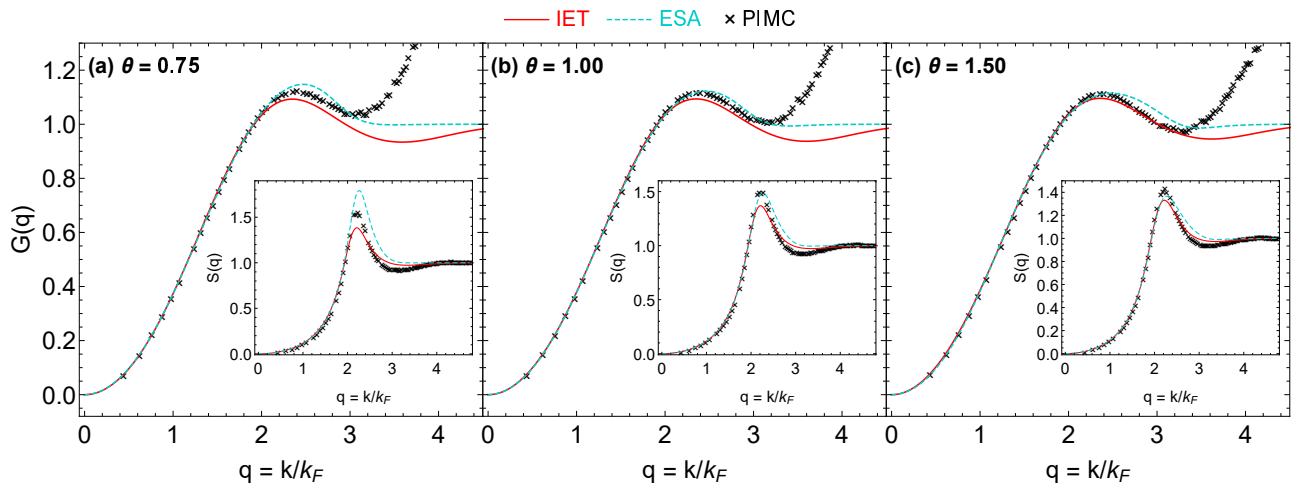}
	\caption{The paramagnetic electron liquid's static local field correction (main) and static structure factor (inset) for $r_{\mathrm{s}}=125$ and $\theta=0.75$ (a), $\theta=1.00$ (b), $\theta=1.50$ (c). Results from the IET scheme (red solid line), the ESA scheme (cyan dashed line) and PIMC simulations (black crosses) in the normalized wavenumber range $k\leq4.5{k}_{\mathrm{F}}$. The ESA shortcoming that concerns the abrupt LFC (and thus SSF) transition from the peak vicinity to the short wavelength limit is apparent.}\label{fig:ESAexamples}
\end{figure*}

Systematic comparisons revealed the following: \textbf{(i)} As expected, the ESA LFC always extrapolates very well towards the long and the short wavelength ranges. \textbf{(ii)} At least for $0.75\leq\theta\leq2.00$ and regardless of $r_{\mathrm{s}}$, the ESA well describes the position and the magnitude of the LFC peak. On the other hand, the ESA strongly overestimates the magnitude of the LFC peak for $\theta=0.5$ especially at strong coupling and underestimates the magnitude of the LFC peak for $\theta=4.0$. \textbf{(iii)} Regardless of the state point of interest, the main drawback of the ESA LFC lies at its transition to the short wavelength regime. This transition is always too abrupt and is not accompanied by small oscillations, as one would expect from a frequency independent LFC at strong coupling (see the IET LFC and the HNC LFC). These characteristics are not present in the warm dense matter regime or in dielectric schemes that are only equipped to describe weak correlations (see the qSTLS LFC). This drawback can be traced back to the implementation of a sigmoid activation function during the construction of ESA\,\cite{ESAsPRB}. \textbf{(iv)} Within $1\leq\theta\leq2$ and regardless of $r_{\mathrm{s}}$, the ESA accurately estimates the position and magnitude of the SSF peak. On the other hand, the ESA strongly overestimates the magnitude of the SSF peak for $\theta=0.5,\,0.75$ especially at strong coupling and underestimates the magnitude of the SSF peak for $\theta=4.0$. \textbf{(v)} Regardless of state point, the ESA SSF directly transitions to its unity short wavelength limit after the Lorentzian peak and does not feature a shallow minimum followed by a second maximum. \textbf{(vi)} For most of the state points, the ESA SSF overestimates the PIMC SSF not only prior to the peak, but also after. As a consequence, it leads to very inaccurate interaction energies. To sum up, the ESA does not extrapolate well to strong coupling, which can mostly be attributed to the abrupt transition from the LFC peak to the short wavelength range. Thus, it can be concluded that the future utilization of an ESA-like scheme at low densities would require a more involved activation function. Some characteristic examples are illustrated in Fig.\ref{fig:ESAexamples}.

\section{Summary and discussion}\label{discussion}

\noindent In this work, we investigated the thermodynamic and the structural properties of the paramagnetic electron liquid with different dielectric theories and with \emph{ab initio} path integral Monte Carlo simulations. First and foremost, we formulated a novel IET scheme that combines the dielectric formalism of many fermion systems with the integral equation theory of classical liquids. Essentially, the IET scheme incorporates a near-exact recent parametrization of the classical OCP bridge function to the existing HNC scheme. In addition, we carried out extensive PIMC simulations of the paramagnetic electron liquid for $16$ state points. Combined with the recent availability of PIMC simulations for $20$ state points, these new results substantially extend our current picture of the finite temperature UEF in the strongly coupled regime.

After an extensive comparison with \enquote{exact} PIMC results, various dielectric formalism schemes and numerous extrapolated high-density parametrizations, we have demonstrated that; \textbf{(i)} The IET scheme yields more accurate results for the interaction energy, static structure factor, static local field correction and static density response of the paramagnetic electron liquid than all other known schemes (HNC, qSTLS, VS, STLS). \textbf{(ii)} The IET interaction energies exhibit a remarkable agreement with the PIMC interaction energies having an accuracy within $0.68\%$ and an average accuracy of $0.29\%$. This has been attributed to a favorable error cancellation in the course of the static structure factor integration. \textbf{(iii)} The IET long wavelength static local field correction, that is connected with the isothermal compressibility through the eponymous sum rule, is nearly indistinguishable from its \enquote{exact} PIMC counterpart up to nearly $k\lesssim2{k}_{\mathrm{F}}$. \textbf{(iv)} The utilization of the IIT interaction energy parametrization leads to very accurate results for the paramagnetic electron liquid (similar accuracy level to the IET), in contrast to extrapolations of the GDSMFB, KSDT and corrKSDT exchange-correlation free energy parametrizations. This has been attributed to the incorporation of exact ground-state and classical limit results as well as to the introduction of an accurate ad hoc $\theta-$interpolation function. \textbf{(v)} The static approximation is much less accurate in the strongly coupled regime than the warm dense matter regime. This confirms that the dynamic nature of the local field correction is an essential ingredient for the paramagnetic electron liquid. On the other hand, the analytic effective static approximation, that is very reliable in the warm dense matter regime, cannot be extrapolated to the strongly coupled regime primarily due to its direct transition to the short wavelength limit after the peak of the frequency independent local field correction.

At this point, it is important to bring forth the two main drawbacks of the IET scheme. \emph{First}, the classical OCP bridge function is introduced as an analytic $b(r,\Gamma)$ parametrization and not as a $b[h]$ functional. As a consequence, the bridge function only reacts to the classical Coulomb interactions and not to quantum mechanical interactions (exchange degeneracy and diffraction effects). This shortcoming manifests itself in the lack of an appropriate ground-state limit given the $\Gamma=2\lambda^2(r_{\mathrm{s}}/\theta)$ mapping of quantum to classical states that stems from the classical coupling parameter definition $\Gamma=e^2/(dk_{\mathrm{B}}T)$. A remedy can be found by recalling the ground-state quantum coupling parameter definition $\Gamma_{\mathrm{q}}=e^2/(dE_{\mathrm{F}})$ that leads to $\Gamma_{\mathrm{q}}=2\lambda^2r_{\mathrm{s}}$. This suggests an effective coupling parameter definition $\Gamma_{\mathrm{eff}}=e^2/\{d[(k_{\mathrm{B}}T)^2+E^2_{\mathrm{F}}]^{1/2}\}$\,\cite{BoniRev} that leads to the new $\Gamma=2\lambda^2r_{\mathrm{s}}/\sqrt{1+\theta^2}$ mapping. The utilization of the effective mapping in lieu of the classical mapping in the IET scheme led to nearly identical results for the state points investigated here. Nevertheless, this should be further tested for low-density state points near the ground-state ($\theta\to0$). It should be mentioned that different empirical mappings could also be utilized based on enforcing consistency with the compressibility sum rule, based on enforcing consistency with some PIMC results (e.g. the static structure factor peak magnitude) or based on adopting empirical relations for a quantum temperature from the classical mapping method\,\cite{outro01,outro02,outro03}. Such options would probably lead to an improved IET scheme that is even more accurate, but they are not desirable because they add a strong phenomenological element to an otherwise rigorous theoretical approach. \emph{Second}, quantum effects are treated on the random phase approximation level. A more sophisticated quantum treatment can be achieved by essentially combining the qSTLS and IET schemes. Given the studied static structure factors of these schemes, this seems to be a very promising strategy that could improve the peak magnitude prediction of the IET scheme. Unfortunately, it also adds an inconsistent element to the approach because the bridge function incorporation technique is based on the classical fluctuation dissipation theorem for a frequency independent local field correction, while the qSTLS scheme necessarily leads to a dynamic local field correction.

Future work will focus on the exploration of the aforementioned possibilities to remedy the drawbacks of the current version of the IET scheme. Moreover, the characterization of the IET scheme's self-consistency concerning the compressibility sum rule and the application of the IET scheme for different values of the spin polarization parameter will also be pursued in the future.

Finally, benefitting from the recent availability of extensive path integral Monte Carlo simulation results for the finite-temperature $\theta\sim1$ paramagnetic electron fluid (non-ideal gas and liquid), the following practical guidelines can be formulated concerning the optimal dielectric formalism scheme. Within the warm dense matter regime of $r_{\mathrm{s}}\lesssim20$, the semi-empirical ESA constitutes the most accurate scheme. At the moderate coupling $20\lesssim{r}_{\mathrm{s}}\lesssim50$ regime, the first-principle HNC scheme is the most accurate. At the strong coupling regime of $r_{\mathrm{s}}\gtrsim50$, the novel first-principle IET scheme is the most accurate. Naturally, the exact quantum coupling parameter boundaries between these three regimes depend on the specific degenerate parameter value.

\section*{Acknowledgments}

\noindent The present work was partly funded by the Swedish National Space Agency under grant no.\,143/16. The present work was also partly funded by the Center for Advanced Systems Understanding (CASUS), which is financed by Germany's Federal Ministry of Education and Research (BMBF) and by the Saxon Ministry for Science, Culture and Tourism (SMWK) with tax funds on the basis of the budget approved by the Saxon State Parliament. All the PIMC simulations were carried out at the Norddeutscher Verbund f\"ur Hoch- und H\"ochstleistungsrechnen (HLRN) under grant no.\,shp00026 and on a Bull Cluster at the Center for Information Services and High Performance Computing (ZIH) at the Technische Universit\"at Dresden. The IET, HNC, qSTLS schemes were numerically solved on resources provided by the Swedish National Infrastructure for Computing (SNIC) at the NSC (Link{\"o}ping University) that is partially funded by the Swedish Research Council under grant agreement no.\,2018-05973.

\section*{Data Availability}

\noindent The data that support the findings of this study are available within this article and from the corresponding author upon reasonable request.

\end{document}